\documentclass[fleqn,oneside,fontsize=10pt,parskip=half,paper=a4]{scrartcl}

\usepackage[left={2.5cm},right={2cm},top={1.5cm},bottom={1.5cm},bindingoffset={0cm},includeheadfoot]{geometry}

\usepackage[utf8]{inputenc}
\usepackage[T1]{fontenc}

\usepackage{lmodern}
\usepackage[ngerman,english]{babel}

\usepackage{graphicx}

\usepackage{setspace}
\usepackage[pdftoolbar=true,pdfmenubar=true,pdfstartview=FitV,pdfdisplaydoctitle=true,pdfencoding=auto,final]{hyperref}
\usepackage[open,openlevel=1,final]{bookmark}

\hypersetup{
    colorlinks = true,
    allcolors = black,
}

\usepackage{abstract}

\usepackage[normal,nooneline,format=plain,singlelinecheck=off,labelfont={bf}]{caption}

\DeclareCaptionFont{sm}{}
\usepackage{subcaption}

\usepackage{floatrow}

\usepackage{csquotes}

\usepackage[backend=biber,style=numeric,uniquename=init,hyperref=true,isbn=false,giveninits=true,sorting=nyt]{biblatex}

\AtEveryBibitem{\iffieldundef{journaltitle}{}{\clearfield{eprint}}}
\AtEveryCitekey{\iffieldundef{journaltitle}{}{\clearfield{eprint}}}

\AtEveryBibitem{%
    \clearfield{issue}%
    \clearfield{number}%
}
\AtEveryCitekey{%
    \clearfield{issue}%
    \clearfield{number}%
}

\DeclareFieldFormat[article]{volume}{\textbf{#1}\setunit*{\space}}
\DeclareFieldFormat[article]{pages}{#1}

\renewcommand*{\bibeidpunct}{\addcolon}

\renewbibmacro*{volume+number+eid}{%
    \printfield{volume}%
    \setunit*{\bibeidpunct}%
    \printfield{eid}%
    \setunit{\addcomma\space}%
}

\renewbibmacro{in:}{
    \ifentrytype{article}{}{\printtext{\bibstring{in}\intitlepunct}}%
}

\renewbibmacro*{journal}{%
    \iffieldundef{shortjournal}%
        {%
            \iffieldundef{journaltitle}%
            {}%
            {\printtext[journaltitle]%
                {%
                    \printfield[titlecase]{journaltitle}%
                    \setunit{\subtitlepunct}%
                    \printfield[titlecase]{journalsubtitle}%
                }%
            }%
        }%
        {%
            \printtext[journaltitle]%
                {%
                    \printfield[titlecase]{shortjournal}%
                }%
        }%
}

\DeclareCiteCommand{\fullcite}
{\usebibmacro{prenote}}
{\usedriver
	{\defcounter{maxnames}{12}}
	{\thefield{entrytype}}}
{\multicitedelim}
{\usebibmacro{postnote}}

\usepackage[noblocks]{authblk}

\usepackage{orcidlink}

\newcommand{\appendixfigures}{\setcounter{figure}{0}\renewcommand{\thefigure}{A\arabic{figure}}}

\usepackage{xcolor}
\usepackage{pdfpages}

\usepackage{tikz}
\usetikzlibrary{decorations.pathreplacing,
decorations.pathmorphing,
calligraphy,
positioning,
quotes,
graphs,
  arrows.meta,
  scopes,
  backgrounds,
  trees,
  calc,
  shapes.multipart
  }
  
  \usepackage{pgfplots}
\pgfplotsset{compat=1.18}

\usepackage{helvet}
\tikzset{every picture/.style={/utils/exec={\sffamily}}}

\usepackage{tabularx}

\definecolor{size-two-hyperedge}{rgb}{0.031,0.28,0.40}
\definecolor{size-three-hyperedge}{rgb}{0.54,0.32,0.55}
\definecolor{size-four-hyperedge}{rgb}{0.98,0.60,0.25}

\makeatletter
\tikzset{circle split part fill/.style  args={#1,#2}{%
 alias=tmp@name,
  postaction={%
    insert path={
     \pgfextra{%
     \pgfpointdiff{\pgfpointanchor{\pgf@node@name}{center}}%
                  {\pgfpointanchor{\pgf@node@name}{east}}%
     \pgfmathsetmacro\insiderad{\pgf@x}
      \fill[#1] (\pgf@node@name.base) 
        ([xshift=-\pgflinewidth]\pgf@node@name.east) arc
            (0:180:\insiderad-\pgflinewidth)--cycle;
      \fill[#2] (\pgf@node@name.base)
        ([xshift=\pgflinewidth]\pgf@node@name.west)  arc
            (180:360:\insiderad-\pgflinewidth)--cycle;
         }
    }
  }
}}  
\makeatother  

\tikzset{%
  A node/.style = {
      draw = black,
      fill = white,
      scale = 0.7,
      shape=circle
    },
  B node/.style = {
      draw = black,
      fill = black!80,
      text = white,
      scale = 0.7,
      shape=circle
    },
  X node/.style = {
    draw = black,
    line width = 0.2mm,
    shape = circle split,
    scale = 0.4,
    circle split part fill={black, white},
    rotate = 225,
  },
  hyperedge/.style = {
      color = #1!30,
      fill = #1!30,
      line width = 8mm,
      line cap = round,
      line join = round,
      preaction={
          draw,
          line width = 9mm,
          color = #1,
          line cap = round,
          line join = round
        }
    },
  hyperedge L/.style = {
      color = #1!30,
      fill = #1!30,
      line width = 11mm,
      line cap = round,
      line join = round,
      preaction={
          draw,
          line width = 12mm,
          color = #1,
          line cap = round,
          line join = round
        }
    },
  hyperedge S/.style = {
      color = #1!30,
      fill = #1!30,
      line width = 5.5mm,
      line cap = round,
      line join = round,
      preaction={
          draw,
          line width = 6.5mm,
          color = #1,
          line cap = round,
          line join = round
        }
    },
  hyperedge XS/.style = {
      color = #1!30,
      fill = #1!30,
      line width = 3.5mm,
      line cap = round,
      line join = round,
      preaction={
          draw,
          line width = 4mm,
          color = #1,
          line cap = round,
          line join = round
        }
    }
}

\newcommand{\sizetwohyperedge}[4]{%
    \node[#1 node] (1) at ($#3 + (0, 0.3)$) {};
    \node[#2 node] (2) at ($#3 - (0, 0.3)$) {};

    \begin{scope}[on background layer]
        \begin{scope}[blend mode = multiply]
            \begin{scope}[transparency group]
                \filldraw [hyperedge #4=size-two-hyperedge] (1.center) -- (2.center);
            \end{scope}
        \end{scope}
    \end{scope}
}

\newcommand{\sizethreehyperedge}[5]{%
    \node[#1 node] (1) at ($#4 + (-0.3, 0.6)$) {};
    \node[#2 node] (2) at ($#4 + (0.3, 0)$) {};
    \node[#3 node] (3) at ($#4 + (-0.3, -0.6)$) {};

    \begin{scope}[on background layer]
        \begin{scope}[blend mode = multiply]
            \begin{scope}[transparency group]
                \filldraw [hyperedge #5=size-three-hyperedge] (1.center) -- (2.center) -- (3.center) -- (1.center);
            \end{scope}
        \end{scope}
    \end{scope}
}

\newcommand{\sizefourhyperedge}[6]{%
    \node[#1 node] (1) at ($#5 + (0, 0.6)$) {};
    \node[#2 node] (2) at ($#5 + (0.6, 0)$) {};
    \node[#3 node] (3) at ($#5 + (0, -0.6)$) {};
    \node[#4 node] (4) at ($#5 + (-0.6, 0)$) {};

    \begin{scope}[on background layer]
        \begin{scope}[blend mode = multiply]
            \begin{scope}[transparency group]
                \filldraw [hyperedge #6=size-four-hyperedge] (1.center) -- (2.center) -- (3.center) -- (4.center) -- (1.center);
            \end{scope}
        \end{scope}
    \end{scope}
}

\usepackage{array}

\usepackage{booktabs}

\usepackage{enumitem}

\usepackage{siunitx}
\KOMAoptions{headings=standardclasses}

\usepackage{multicol}

\newcommand{\twocolumngeometry}{\newgeometry{left={1.4cm},right={1.4cm},top={0.5cm},bottom={1cm},includeheadfoot,footskip={1cm}}} 
\newcommand{\openwidetext}{\end{multicols}\rule{\dimexpr(0.5\textwidth-0.5\columnsep-0.4pt)}{0.4pt}\rule{0.4pt}{6pt}}
\newcommand{\closewidetext}{\hfill\rule[-6pt]{0.4pt}{6.4pt}\rule{\dimexpr(0.5\textwidth-0.5\columnsep-1pt)}{0.4pt}\begin{multicols}{2}}
	
\newcommand{\openwidefigure}{\end{multicols}}
\newcommand{\closewidefigure}{\begin{multicols}{2}}

\usepackage{amsmath}
\usepackage{amsfonts}
\usepackage{amssymb}
\usepackage{amsthm}

\DeclareSymbolFont{dsrom}{U}{dsrom}{m}{n}
\DeclareMathSymbol{\dsI}{3}{dsrom}{"31}
\DeclareSymbolFont{bbold}{U}{bbold}{m}{n}
\DeclareSymbolFontAlphabet{\mathbbold}{bbold}

\newcommand{\mathrelphantom}[1]{\ensuremath{\mathrel{\phantom{#1}}}}

\newcommand{\naturals}{\ensuremath{\mathbb{N}}}

\newcommand{\indicator}[1]{\ensuremath{\dsI_{#1}}}

\newcommand{\stack}[2]{\ensuremath{\genfrac{}{}{0pt}{}{#1}{#2}}}
\newcommand{\triplestack}[3]{\ensuremath{\stack{#1}{\stack{#2}{#3}}}}

\newcommand{\derivative}[2][]{\ensuremath{\left.\tfrac{\mathrm{d}}{\mathrm{d}#2}\ifthenelse{\isempty{#1}}{\right.}{\right\vert_{#2=#1}}}}
\newcommand{\tderivative}[2][]{\ensuremath{\tfrac{\mathrm{d}}{\mathrm{d}#2}\ifthenelse{\isempty{#1}}{}{\vert_{#2=#1}}}}

\renewcommand{\emptyset}{\ensuremath{\varnothing}}

\newcommand{\tanglebr}[1]{\ensuremath{\langle #1 \rangle}}

\newcommand{\A}{\ensuremath{\mathrm{A}}}
\newcommand{\B}{\ensuremath{\mathrm{B}}}
\newcommand{\X}{\ensuremath{\mathrm{X}}}

\newcommand{\sumactiveedges}{\sum_{\stack{h \in \{\A^{m}\B^{n}\}}{(m,n) \in \bar{Q}^{K}}}}

\newcommand{\sumsubsetsAplus}{\sum_{\triplestack{\alpha' \subseteq \alpha, \emptyset \neq \beta' \subseteq \beta}{|\alpha'| + |\beta'| < \max(a+b, m+n)}{|\alpha'| + |\beta'| \leq a}}}
\newcommand{\sumsubsetsAminus}{\sum_{\triplestack{\alpha' \subseteq \alpha, \emptyset \neq \beta' \subseteq \beta}{|\alpha'| + |\beta'| < \max(a+b, m+n)}{|\alpha'| \leq a, |\beta'| \leq b}}}
\newcommand{\sumsubsetsBplus}{\sum_{\triplestack{\emptyset \neq \alpha' \subseteq \alpha, \beta' \subseteq \beta}{|\alpha'| + |\beta'| < \max(a+b, m+n)}{|\alpha'| + |\beta'| \leq b}}}
\newcommand{\sumsubsetsBminus}{\sum_{\triplestack{\emptyset \neq \alpha' \subseteq \alpha, \beta' \subseteq \beta}{|\alpha'| + |\beta'| < \max(a+b, m+n)}{|\alpha'| \leq a, |\beta'| \leq b}}}

\newcommand{\sumintersectionsAplus}{\sum_{\triplestack{0 \leq \mu \leq m, 1 \leq \nu \leq n}{\mu + \nu < \max(m+n, a+b)}{\mu + \nu \leq a}}}
\newcommand{\sumintersectionsAminus}{\sum_{\triplestack{0 \leq \mu \leq m, 1 \leq \nu \leq n}{\mu + \nu < \max(m+n, a+b)}{\mu \leq a, \nu \leq b}}}
\newcommand{\sumintersectionsBplus}{\sum_{\triplestack{1 \leq \mu \leq m, 0 \leq \nu \leq n}{\mu + \nu < \max(m+n, a+b)}{\mu + \nu \leq b}}}
\newcommand{\sumintersectionsBminus}{\sum_{\triplestack{1 \leq \mu \leq m, 0 \leq \nu \leq n}{\mu + \nu < \max(m+n, a+b)}{\mu \leq a, \nu \leq b}}}

\usepackage{xifthen}

\usepackage{xstring}

\RequirePackage[en-US]{datetime2}
\DTMlangsetup[en-US]{ord=raise}

\usepackage{scrhack}

\usepackage[final]{microtype}

\title{Polyadic Opinion Formation: The Adaptive Voter Model on a Hypergraph}

\author[1,2]{Anastasia Golovin~\orcidlink{0009-0005-3490-1354}\,}
\author[1]{Jan Mölter~\orcidlink{0000-0002-5964-6207}\,}
\author[1,3,4]{Christian Kuehn~\orcidlink{0000-0002-7063-6173}\,}

\affil[1]{Department of Mathematics, School of Computation, Information and Technology, Technical University of Munich, Boltzmannstraße 3, 85748 Garching bei München, Germany}
\affil[2]{Max-Planck-Institute for Dynamics and Self-Organization, Am Faßberg 17, 37077 Göttingen, Germany}
\affil[3]{Munich Data Science Institute, Technical University of Munich, Walther-von-Dyck-Straße 10, 85748 Garching bei München, Germany}
\affil[4]{Complexity Science Hub Vienna, Josefstädter Straße 39, 1080 Vienna, Austria}

\date{}

\begin{document}

\maketitle
\setcounter{page}{1}

\begin{abstract}
    The adaptive voter model is widely used to model opinion dynamics in social complex networks. However, existing adaptive voter models are limited to only pairwise interactions and fail to capture the intricate social dynamics that arises in groups. This paper extends the adaptive voter model to hypergraphs to explore how forces of peer pressure influence collective decision-making. The model consists of two processes: individuals can either consult the group and change their opinion or leave the group and join a different one. The interplay between those two processes gives rise to a two-phase dynamics. In the initial phase, the topology of the hypergraph quickly reaches a new stable state. In the subsequent phase, opinion dynamics plays out on the new topology depending on the mechanism by which opinions spread. If the group always follows the majority, the network rapidly converges to fragmented communities. In contrast, if individuals choose an opinion proportionally to its representation in the group, the system remains in a metastable state for an extended period of time. The results are supported both by stochastic simulations and an analytical mean-field description in terms of hypergraph moments with a moment closure at the pair level. 

    \vspace{2\parsep}

    \textbf{Keywords} {higher-order interactions} $\cdot$ {adaptive voter models} $\cdot$ {mean-field models} $\cdot$ {moment closures} $\cdot$ {phase transitions}
\end{abstract}

\setcounter{secnumdepth}{0}
\tableofcontents

\twocolumngeometry

\begin{multicols}{2}

\section{Introduction}

Opinions of individuals in a society are not formed in isolation but are rather shaped through communication and interaction with others. Such interactions happen not only in personal face-to-face conversations but also in larger groups, where forces of conformity and peer pressure may affect the outcome of a discussion~\parencite{asch1961effects,crandall2002social,brown1986perceptions,hansen1991preventing}. Consequently, models of opinion dynamics must take group interactions into account.

A classical model to study opinion and consensus formation is the voter model~\parencite{clifford1973model,holley1975ergodic,liggett1999stochastic}. Individuals in this model are represented as vertices in a network and are assigned one of two opposing opinions. The individuals change their opinions at random times sampled from a Poisson process by copying the opinion of one of their neighbors. By this process, the system evolves until it either reaches consensus (all individuals hold the same opinion) or a state in which both opinions coexist. One of the key findings in this model is that the topology of the network strongly influences whether consensus can be achieved or not~\parencite{holley1975ergodic,frachebourg1996exact,suchecki2004conservation,sood2005voter}.

Realistically, the social network utilized within a voter model is not static; instead, individuals may seek to surround themselves with like-minded people~\parencite{mcpherson2001birds}. To include this effect, the voter model was extended with an adaptive component leading to the adaptive voter model~\parencite{holme2006nonequilibrium,gross2008adaptive,benczik2009opinion,zschaler2012adaptive}. Compared to the classical voter model, two kinds of interactions are possible in the adaptive variant: With probability $1 - p$, an individual copies the opinion of a neighbor, with probability $p$, they break up the connection and connect with a different individual. This interplay between opinion propagation and network adaptation gives rise to a new phase transition at a critical value $p = p_{c}$~\parencite{holme2006nonequilibrium,kimura2008coevolutionary,vazquez2008generic}. If opinion propagation dominates ($p < p_{c}$), the network remains in a metastable state with a non-vanishing density of connections between individuals with different opinions. In contrast, if adaptation dominates ($p > p_{c}$), the network fragments into disconnected communities, in each of which consensus is reached. As such, the phase transition between these two states is called the fragmentation transition.

\begin{figure}[H]
	\centering
	\includegraphics{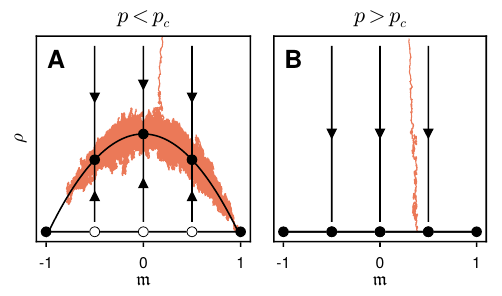}
	\caption{A sketch of the phase space of the adaptive voter model before (panel \textbf{A}) and after (panel \textbf{B}) the fragmenting phase transition. The orange line shows a typical stochastic trajectory.}
	\label{fig:phase_space_sketch}
\end{figure}

It was shown that the mechanism behind the phase transition can be understood by plotting the evolution of the stochastic system in the $(\mathfrak{m}, \rho)$ coordinates, where $\mathfrak{m}$ is the magnetization of the system defined as the difference between the density of both opinions, and $\rho$ is the density of edges between vertices with different opinions, the so-called active edges~\parencite{kimura2008coevolutionary,vazquez2008generic}. For $p < p_{c}$, this plot reveals a parabola-shaped curve. Analytical mean-field calculations show that the parabola corresponds to a slow manifold of critical points with one attractive and one degenerate direction (cf. Fig.~\ref{fig:phase_space_sketch})~\parencite{kimura2008coevolutionary}. Once the system reaches a point on the parabola, it stays in this metastable state for a long time. On finite networks, stochastic fluctuations eventually bring the system to one of the absorbing states at the roots of the parabola, so the parabola can also be interpreted as a slow manifold. If the adaptation probability $p$ is increased, the height of the parabola decreases and reaches zero at the critical point $p_{c}$. For $p > p_{c}$, the system rapidly depletes all active edges and converges to a fragmented network in which individuals with different opinions do not interact with each other.

Although the adaptive voter model is more realistic compared to the classical non-adaptive variant, it neglects the impact of group interactions on the opinion formation. In recent years, there has been an increasing amount of research on interactions of higher order in networks, i.e., interactions which involve multiple vertices~\parencite{battiston2020networks,majhi2022dynamics}. Such models generally use either simplicial complexes~\parencite{hatcher2003algebraic} or hypergraphs~\parencite{bretto2013hypergraph} as a generalization of networks~\parencite{battiston2020networks}. The interactions of higher order give rise to new interesting phenomena. For example, in models of epidemic or rumor spreading~\parencite{iacopini2019simplicial} and for coupled oscillators on simplicial complexes~\parencite{skardal2019abrupt}, it was observed that the higher-order interactions lead to a discontinuous phase transition~\parencite{kuehn2021universal}. The continuous-to-discontinuous change of the oscillator phase transition has recently been proven mathematically for hypergraph mean-fields~\parencite{bick2023higher}.

With regard to opinion formation, one of the earliest works considered the adaptive voter model on low-dimensional simplicial complexes~\parencite{horstmeyer2020adaptive}. The authors showed that group interactions accelerate the convergence of the system to an absorbing state. This observation has been confirmed in a recent work that extended the adaptive voter model to hypergraphs~\parencite{papanikolaou2022consensus}.

In this work, we continue the exploration of the adaptive voter model on hypergraphs. We consider four evolution dynamics that are composed of two different adaptation, as well as two different propagation rules.
We derive a corresponding mean-field description in terms of hypergraph moments that extends previous works and shows overall good agreement with extensive simulations of the stochastic particle model.

Moreover, we find that if the individuals do not always follow the majority opinion but instead choose a new opinion proportionally to its representation in the group, the system no longer converges as rapidly to an absorbing state as reported before~\parencite{horstmeyer2020adaptive,papanikolaou2022consensus}, but instead remains in a metastable state. This allows us to investigate the interplay between the propagation of opinions and adaptive rewiring. Specifically, we show that the topology of the hypergraph is first shaped on a fast timescale; after that, the system evolves slowly until it eventually reaches an absorbing state. In addition, we also report the occurrence of a fragmentation transition as in earlier works, yet with the critical point shifted. Finally, these findings, the rapid convergence of the hypergraph to a stable topology, the slow diffusion of states, and the phase transition, are also qualitatively reproduced and observed in the mean-field.

\section{Results}

\subsection{The Adaptive Voter Model on a Hypergraph}

The classical adaptive voter model considers a graph in which every vertex is assigned one of two possible opinions, $\A$ or $\B$~\parencite{holme2006nonequilibrium,demirel2014momentclosure,zschaler2012adaptive}. In the link-update variant of the model, the dynamics is initiated by active edges, i.e., edges that connect vertices with different opinions, rather than the vertices themselves~\parencite{demirel2014momentclosure}. Specifically, active edges trigger interactions subject to a Poisson process at a constant rate that can be rescaled to one. When an interaction occurs, a vertex chosen uniformly at random either breaks the edge and rewires it to a different vertex with probability $p$ (\enquote{adaptation}), or copies its neighbor's opinion with probability $1 - p$ (\enquote{propagation}).

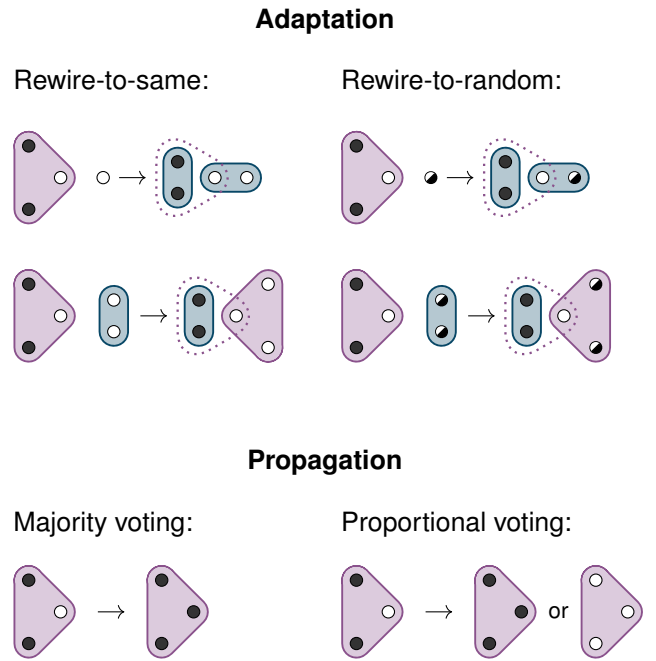
\begin{figure}[H]
	\sffamily

\begin{tikzpicture}
    \begin{scope}[shift={(0, 0)}]
        \node[align=center, anchor=center] at (4, 1) {\textbf{Adaptation}};
        \node (leftcol) at (0, 0){};
        \node (rightcol) at (4.25, 0){};
        \node[align=left, anchor=west] at ($(leftcol)$) {Rewire-to-same:};
        \begin{scope}[baseline=-0.65ex, scale=0.7, every node/.style={transform shape}, shift={($(leftcol) + (0.8, -1.7)$)}]
            \sizethreehyperedge{B}{A}{B}{(0, 0)}{XS}
            \node[A node] () at (1.1, 0) {};
            \draw[->] (1.4, 0) -- (1.9, 0);
            \sizetwohyperedge{B}{B}{(2.5, 0)}{XS}
            \begin{scope}[rotate around={90:(3.5, 0)}]
                \sizetwohyperedge{A}{A}{(3.5, 0)}{XS}
            \end{scope}
            \draw[dotted, rounded corners=5mm,color=size-three-hyperedge,line width=0.3mm] (2.1, 1)--(3.8, 0)--(2.1, -1)--cycle;
        \end{scope}
        \begin{scope}[baseline=-0.65ex, scale=0.7, every node/.style={transform shape}, shift={($(leftcol) + (0.8, -4.3)$)}]
            \sizethreehyperedge{B}{A}{B}{(0, 0)}{XS}
            \sizetwohyperedge{A}{A}{(1.3, 0)}{XS}
            \draw[->] (1.8, 0) -- (2.3, 0);
            \sizetwohyperedge{B}{B}{(2.9, 0)}{XS}
            \begin{scope}[rotate around={180:(3.9, 0)}]
                \sizethreehyperedge{A}{A}{A}{(3.9, 0)}{XS}
            \end{scope}
            \draw[dotted, rounded corners=5mm,color=size-three-hyperedge,line width=0.3mm] (2.5, 1)--(4.2, 0)--(2.5, -1)--cycle;
        \end{scope}
        \node[align=left, anchor=west] at ($(rightcol)$) {Rewire-to-random:};
        \begin{scope}[baseline=-0.65ex, scale=0.7, every node/.style={transform shape}, shift={($(rightcol) + (0.8, -1.7)$)}]
            \sizethreehyperedge{B}{A}{B}{(0, 0)}{XS}
            \node[X node] (1) at (1.1, 0) {};
            \draw[->] (1.4, 0) -- (1.9, 0);
            \sizetwohyperedge{B}{B}{(2.5, 0)}{XS}

            \node[A node] (2) at (3.2, 0) {};
            \node[X node] (3) at (3.8, 00) {};
            \begin{scope}[on background layer]
                \begin{scope}[blend mode = multiply]
                    \begin{scope}[transparency group]
                        \filldraw [hyperedge XS=size-two-hyperedge] (2.center) -- (3.center);
                    \end{scope}
                \end{scope}
            \end{scope}

            \draw[dotted, rounded corners=5mm,color=size-three-hyperedge,line width=0.3mm] (2.1, 1)--(3.8, 0)--(2.1, -1)--cycle;
        \end{scope}
        \begin{scope}[baseline=-0.65ex, scale=0.7, every node/.style={transform shape}, shift={($(rightcol) + (0.8, -4.3)$)}]
            \sizethreehyperedge{B}{A}{B}{(0, 0)}{XS}
            \sizetwohyperedge{X}{X}{(1.3, 0)}{XS}
            \draw[->] (1.8, 0) -- (2.3, 0);
            \sizetwohyperedge{B}{B}{(2.9, 0)}{XS}

            \node[A node] (2) at (3.6, 0) {};
            \node[X node] (3) at (4.2, 0.6) {};
            \node[X node] (4) at (4.2, -0.6) {};
            \begin{scope}[on background layer]
                \begin{scope}[blend mode = multiply]
                    \begin{scope}[transparency group]
                        \filldraw [hyperedge XS=size-three-hyperedge] (2.center) -- (3.center) -- (4.center) -- cycle;
                    \end{scope}
                \end{scope}
            \end{scope}
            \draw[dotted, rounded corners=5mm,color=size-three-hyperedge,line width=0.3mm] (2.5, 1)--(4.2, 0)--(2.5, -1)--cycle;
        \end{scope}
    \end{scope}
    \begin{scope}[shift={(0, -6)}]
        \node[align=center, anchor=center] at (4, 1) {\textbf{Propagation}};
        \node (leftcol) at (0, 0){};
        \node (rightcol) at (4.25, 0){};
        \node[align=left, anchor=west] at ($(leftcol)$) {Majority voting:};
        \begin{scope}[baseline=-0.65ex, scale=0.7, every node/.style={transform shape}, shift={(0.8, -1.7)}]
            \sizethreehyperedge{B}{A}{B}{(0, 0)}{XS}
            \draw[->] (1.0, 0) -- (1.5, 0);
            \sizethreehyperedge{B}{B}{B}{(2.5, 0)}{XS}
        \end{scope}
        \node[align=left, anchor=west] at ($(rightcol)$) {Proportional voting:};
        \node (bottomright) at (8.4, -2){};
        \begin{scope}[baseline=-0.65ex, scale=0.7, every node/.style={transform shape}, shift={($(rightcol) + (0.8, -1.7)$)}]
            \sizethreehyperedge{B}{A}{B}{(0, 0)}{XS}
            \draw[->] (1.0, 0) -- (1.5, 0);
            \sizethreehyperedge{B}{B}{B}{(2.5, 0)}{XS}
            \node[] () at (3.5, 0) {\large{or}};
            \sizethreehyperedge{A}{A}{A}{(4.5, 0)}{XS}
        \end{scope}
    \end{scope}
\end{tikzpicture}
	\caption{Four variants of update rules for an example active hyperedge with one $\A$-vertex (white) and two $\B$-vertices (black). During an adaptation event, a random vertex leaves the hyperedge, in this case, the $\A$-vertex. Under the rewire-to-same rule, it can join only other $\A$-vertices or hyperedges which consist only of $\A$-vertices; under the rewire-to-random rule, it can join any vertex or hyperedge. During a propagation event, all vertices in the hyperedge switch to the same opinion. Under majority voting, the majority opinion $\B$ is always chosen, under proportional voting, opinion $\A$ is chosen with probability $1/3$ and opinion $\B$ with probability $2/3$.}
	\label{fig:rules_illustration}
\end{figure}

The hypergraph model follows the same general idea. We start with a random hypergraph with $N$ vertices and a fixed number of hyperedges of every cardinality, $\mathbf{M} = (M_{2}, M_{3}, \hdots,  M_{K})$, up to the maximum cardinality $K$. Hyperedges that contain only one vertex are forbidden to ensure that the hypergraph reduces to a graph for $K = 2$. Similarly to the graph model, interactions occur in active hyperedges, i.e., hyperedges that contain vertices of different opinions, subject to a Poisson process. Each interaction can be either an adaptation event with probability $p$ or a propagation event with probability $1 - p$. For both events, we will consider two variants of the update rules summarized in Fig.~\ref{fig:rules_illustration}.

\paragraph{Adaptation} During an adaptation event, a vertex selected uniformly at random among the ones of the hyperedge leaves it and joins a vertex or an existing hyperedge of cardinality less than $K$. This target vertex or hyperedge is chosen uniformly at random from the set of all admissible vertices and hyperedges. If the source hyperedge contained only two vertices, it is destroyed by this operation, otherwise, it continues to exist as a smaller hyperedge. Similarly, if the free vertex joins another vertex, a new hyperedge of cardinality two is created, but if it joins an existing hyperedge, this hyperedge increases in size. Note that for general $N$ and $\mathbf{M}$ and for $K > 2$, this rewiring process does not conserve the total number of hyperedges or the number of hyperedges of a particular cardinality.

In models on graphs, one distinguishes between two variants of adaptation dynamics known as the rewire-to-same and rewire-to-random rules~\parencite{durrett2012graph,chodrow2020local}. The rewire-to-same rule only allows a vertex to join a vertex with the same opinion, while the rewire-to-random rule lets a vertex join any other vertex regardless of its opinion. To generalize those rules to hypergraphs, we will require that under the rewire-to-same rule, \emph{all} vertices in the target hyperedge have to share the same opinion as the joining vertex. On the other hand, the generalized rewire-to-random rule does not place any restrictions on the target vertex or hyperedge.

\paragraph{Propagation} When a propagation event occurs, all vertices in the hyperedge switch to the same opinion. Two options here are to choose either the opinion of an arbitrary vertex, selected uniformly at random, or the opinion of the majority. We refer to these as the proportional voting and majority voting variants. While in the latter, the majority's opinion is always propagated (in case of a tie, an opinion is chosen randomly), in the former, either opinion is propagated with a probability proportional to its prevalence in the hyperedge.

\subsection{Derivation of a Mean-Field Description}

\subsubsection{Notation}
In the following, we derive a mean-field description in terms of certain hypergraph motifs similar to the models on graphs~\parencite{sharkey2008deterministic,house2009motif,pellis2015exact,kiss2017mathematics}; see also~\parencite{bodo2016sis,schlager2022stability} for the use of motifs in higher-order interaction models of epidemics and game theory. In our context, we let $[\Xi]$ denote the expected number of hyperedges matching the pattern $\Xi$ in the hypergraph. Similarly, we let $[\Xi_{1} (\Xi_{1 \cap 2}) \Xi_{2}]$ denote the expected number of two intersecting hyperedges for which the vertices in the first hyperedge that do not intersect with the second hyperedge match the pattern $\Xi_{1}$, the vertices in which the first and second hyperedge intersect match the pattern $\Xi_{1 \cap 2}$, and the vertices in the second hyperedge that do not intersect with the first hyperedge match $\Xi_{2}$. Finally, we let $\{\Xi\}$ denote the set of all hyperedges matching the pattern $\Xi$.

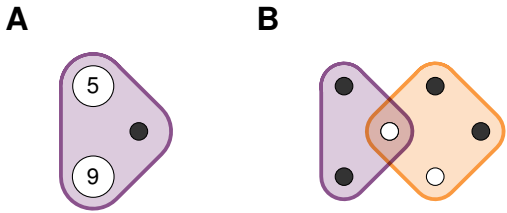
\begin{figure}[H]
	\centering
\begin{minipage}[c][4cm]{1\textwidth}%
    \centering
    \begin{tikzpicture}
        \node (bottomright) at (5, -1) {};

        \node[font=\sffamily] at (-1.6, 1.5) {\large{\textbf{A}}};
        \node[A node, scale=1.2] (1) at (-0.6, 0.6) {5};
        \node[A node, scale=1.2] (2) at (-0.6, -0.6) {9};
        \node[B node] (3) at (0, 0) {};
        \begin{scope}[on background layer]
            \begin{scope}[blend mode = multiply]
                \begin{scope}[transparency group]
                    \filldraw [hyperedge=size-three-hyperedge] (1.center) -- (2.center) -- (3.center) -- (1.center);
                \end{scope}
            \end{scope}
        \end{scope}

        \node[font=\sffamily] at (1.7, 1.5) {\large{\textbf{B}}};
        \sizethreehyperedge{B}{A}{B}{(3, 0)}{S}
        \sizefourhyperedge{B}{B}{A}{A}{(3.9, 0)}{S}
    \end{tikzpicture}
\end{minipage}
	\caption{Examples of hyperedge patterns. $\A$-vertices are white, $\B$-vertices are black. \textbf{A}: Pattern $\A^{2}_{\{5, 9\}} \B$ matches any hyperedge consisting of two $\A$-vertices with indices 5 and 9 and any $\B$-vertex. \textbf{B}: Pattern $\B^{2} ( \A ) \A \B^{2}$ matches any two hyperedges, one with one $\A$-vertex and two $\B$-vertices and the other with two $\A$-vertices and two $\B$-vertices, which share an $\A$-vertex.} \label{fig:notation_example}
\end{figure}

In this context, a pattern will be a sequence of labels describing a certain composition of states. The pattern matching a set of $a$ $\A$-vertices and $b$ $\B$-vertices is written as $\underbrace{\A \cdots \A}_{\text{$a$ times}}\underbrace{\B \cdots \B}_{\text{$b$ times}}$, where the order does not matter, or $\A^{a}\B^{b}$ for short. Occasionally, we need to describe the compositions of states around specific vertices. For that, we will use subscripts to refer to these vertices' indices. This way, the pattern $\A_{i_{1}} \cdots \A_{i_{a'}} \A^{a} \B_{j_{1}} \cdots \B_{j_{b'}} \B^{b}$ or $\A_{\{i_{1}, \ldots i_{a'}\}} \A^{a} \B_{\{j_{1}, \ldots j_{b'}\}} \B^{b}$ for short describes the set of $a + a'$ $\A$-vertices including vertices $i_{1}$, \ldots $i_{a'}$ and $b + b'$ $\B$-vertices including vertices $j_{1}$, \ldots $j_{b'}$ (cf. Fig.~\ref{fig:notation_example}).

Together, this provides the required flexibility to generalize the notation used elsewhere to denote graph motifs to the hypergraph setting~\parencite{kimura2008coevolutionary,demirel2014momentclosure}. In particular, the notation of moments of order zero (a single vertex) and one (a single hyperedge) remains the same, while a moment of order two that corresponds to a triple such as $[ \X\X'\X'' ]$ for $\X, \X', \X'' \in \{\A,\B\}$ is denoted as $[ \X (\X') \X'' ]$ here.

Finally, as events can only be initiated by active hyperedges, we frequently need to iterate over the set of all active hyperedges. For that, we let
\begin{equation}
	\bar{Q}^K := \{(m, n) \in \naturals^2 \mid 0 < m, n \text{ and } m + n \leq K\}
\end{equation}
denote the set of all possible compositions of hyperedges such that $\A^{m} \B^{n}$ is active.

\subsubsection{General Structure of Equations}

The mean-field equations take the general form~\parencite{rand1999correlation}
\begin{equation}
	\derivative{t} [\A^{a}\B^{b}] = \sumactiveedges{} \Delta [\A^{a}\B^{b}]_{(h)},
	\label{eq:general_form}
\end{equation}
for all $1 \leq a + b \leq K$. Here, the sum runs over all active hyperedges, and $\Delta [\A^{a}\B^{b}]_{(h)}$ denotes the change in the expected number of $\A^{a}\B^{b}$ motifs due to any interactions initiated by the active hyperedge $h$.

By the law of total expectation, we have that
\begin{equation}
	\Delta [\A^{a}\B^{b}]_{(h)} = p \Delta [\A^{a}\B^{b}]_{(h, \text{adapt.})} + (1-p) \Delta [\A^{a}\B^{b}]_{(h, \text{prop.})},
	\label{eq:split_expected_value}
\end{equation}
where, analogously to what we had before, $\Delta [\A^{a}\B^{b}]_{(h,\text{adapt.})}$ and $\Delta [\A^{a}\B^{b}]_{(h,\text{prop.})}$ denote the change in the expected number of $\A^{a}\B^{b}$ motifs due to a single adaptation and propagation event, respectively, initiated by an active hyperedge $h$. After inserting this into Eq. \eqref{eq:general_form} and carrying out the sum, we obtain the final form of the mean-field equations
\begin{equation}
	\derivative{t} [\A^{a}\B^{b}] = p \, \Delta [\A^{a}\B^{b}]_{(\text{adapt.})} + (1-p) \, \Delta [\A^{a}\B^{b}]_{(\text{prop.})},
	\label{eq:mean-field-equation}
\end{equation}
where $\sum_{h} \Delta [\A^{a}\B^{b}]_{(h, \text{adapt.})} =: \Delta [\A^{a}\B^{b}]_{(\text{adapt.})}$ is the expected change due to all adaptation events in all active hyperedges and $\sum_h \Delta [\A^{a}\B^{b}]_{(h, \text{prop.})} =: \Delta [\A^{a}\B^{b}]_{(\text{prop.})}$ is the expected change due to all propagation events. The next two sections are fully dedicated to computing those two terms.

\subsubsection{Adaptation Contribution}

Since only the topology of the network changes under adaptation, the expected number of $\A$- and $\B$-vertices remains constant. Hence,
\begin{equation}
	\begin{split}
		\Delta[\A]_{(\text{adapt.})} & = 0  \\
		\Delta[\B]_{(\text{adapt.})} & = 0.
	\end{split}
\end{equation}

\begin{figure}[H]
	\begin{minipage}[c][5cm]{\textwidth}
    \centering
    \begin{tikzpicture}[empty vertex/.style={coordinate,circle,inner sep=0,minimum size=1.5mm}]

        \begin{scope}
            \node[empty vertex, align=center, font=\footnotesize] (start) at (-0.8, 0) {$\A$ leaves\\$h \in \{\A^m \B^n\}$};
            \node[empty vertex] (A_leaves_left) at (0, 0) {};
            \node[empty vertex] (A_leaves_right) at (3.5, 0) {};
            \node[empty vertex] (A_joins_left) at (8, 0) {};
            \node[empty vertex] (A_joins_right) at (11.5, 0) {};

            \draw[->] (A_leaves_left) -- node[above, text width=1.1cm, align=center, yshift=0.2cm] {\scriptsize $\A$ leaves $\A^{a + 1}\B^{b}$} ($(A_leaves_right) + (0, 1)$);
            \draw[->] (A_leaves_left) -- node[below, text width=1.1cm, align=center, yshift=-0.2cm] {\scriptsize $\A$ leaves $\A^{a}\B^{b}$} ($(A_leaves_right) - (0, 1)$);

            \node[fill=black,opacity=0.1,rectangle,anchor=center,minimum width=1.1cm,minimum height=3cm,xshift=1cm] at (A_leaves_right) {};

            \node [above, align=center, font=\footnotesize] at ($(A_leaves_right) + (1, 1.5)$) {change in\\$[\A^a \B^b]$};
            \node at ($(A_leaves_right) + (1, 1)$) {$+1$};
            \node at ($(A_leaves_right) + (1, -1)$) {$-1$};

            \draw [decorate,decoration={calligraphic brace,amplitude=6pt},ultra thick] ($(A_joins_left) + (-0.3, 1.3)$) -- ($(A_joins_left) + (-0.3, -1.3)$);

            \node [anchor=south west, font=\footnotesize] at ($(A_leaves_right) + (2, 1.5)$) {probability};
            \node [right] at ($(A_leaves_right) + (2, 1)$) {$\delta_{m, a + 1}\delta_{n, b}$};
            \node [right] at ($(A_leaves_right) + (2, -1)$) {$\delta_{m, a}\delta_{n, b}$};

            \draw[->] (A_joins_left) -- node[above, text width=1.0cm, align=center, yshift=0.2cm] {\scriptsize $\A$ joins $\A^{a-1}\B^{b}$} ($(A_joins_right) + (0, 1)$);
            \draw[->] (A_joins_left) -- node[below,text width=1.0cm, align=center, yshift=-0.2cm] {\scriptsize $\A$ joins $\A^{a}\B^{b}$} ($(A_joins_right) - (0, 1)$);

            \node[fill=black,opacity=0.1,rectangle,anchor=center,minimum width=1.1cm,minimum height=3cm,xshift=1cm] at (A_joins_right) {};

            \node [above, align=center, font=\footnotesize] at ($(A_joins_right) + (1, 1.5)$) {change in\\$[\A^a \B^b]$};
            \node at ($(A_joins_right) + (1, 1)$) {$+1$};
            \node at ($(A_joins_right) + (1, -1)$) {$-1$};

            \node [anchor=south west, font=\footnotesize] at ($(A_joins_right) + (2, 1.5)$) {probability};
            \node [right] at ($(A_joins_right) + (2, 1)$) {$\pi_{\A}(a - 1, b)$};
            \node [right] at ($(A_joins_right) + (2, -1)$) {$\pi_{\A}(a, b)$};
        \end{scope}
    \end{tikzpicture}
\end{minipage}
	\caption{All possible changes in $[\A^{a} \B^{b}]$ caused by an $\A$-vertex leaving the hyperedge $h \in \{\A^{m} \B^{n}\}$. The corresponding diagram for a $\B$-vertex leaving can be obtained analogously.}
	\label{fig:adaptivity_transitions}
\end{figure}
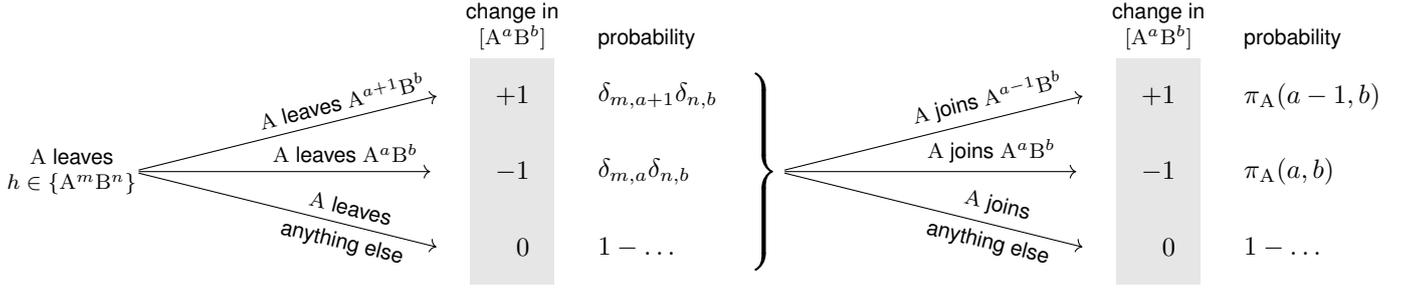

Therefore, we only need to consider the hyperedge motifs $\A^{a} \B^{b}$ with $2 \leq a + b \leq K$. To begin, assume that an adaptation event takes place in an active hyperedge $h \in \{\A^{m}\B^{n}\}$ ($1 \leq m,n$). We want to compute the expected change in $[\A^{a} \B^{b}]$ caused by this event. An $\A^{a} \B^{b}$ hyperedge can be created or destroyed in two ways: first, when a vertex leaves a hyperedge, second, when it joins a new hyperedge or a different vertex. The possible transitions together with the corresponding changes are summarized in Fig.~\ref{fig:adaptivity_transitions}.

\openwidefigure
\closewidefigure

Assume first that the vertex leaving $h$ is in state $\A$. If $m = a$ and $n = b$, in other words, if $h$ already has the form $\A^{a} \B^{b}$, the number of $\A^{a} \B^{b}$ hyperedges decreases by one. On the other hand, if $m = a + 1$ and $n = b$, an $\A^{a} \B^{b}$ hyperedge is created.

Next, the free $\A$-vertex joins either a different vertex or a different hyperedge. Let $\pi_{\X}(a,b)$ denote the probability that a free $\X$-vertex connects with a hyperedge $\A^{a}\B^{b}$ (or a vertex if $(a, b) \in \{(1, 0),\, (0, 1)\}$). In particular, we have that
\begin{itemize}[leftmargin=*]
	\item under the rewire-to-random rule,
		\begin{equation*}
			\begin{split}
				\pi_{\X}(a, b) & \propto
				\begin{cases}
					[\A^{a}\B^{b}] & $if $ 1 \leq a + b < K $ and $ 0 \leq a, b \\
					0              & \text{otherwise,}
				\end{cases}
				\\
			\end{split}
		\end{equation*}
	\item under the rewire-to-same rule, \\
		\begin{equation*}
			\begin{split}
				\pi_{\A}(a, b) & \propto
				\begin{cases}
					[\A^{a}] & \text{if $1 \leq a < K$ and $b = 0$,} \\
					0        & \text{otherwise}
				\end{cases}
				\\
				\pi_{\B}(a, b) & \propto
				\begin{cases}
					[\B^{b}] & \text{if $a = 0$ and $1 \leq b < K$,} \\
					0        & \text{otherwise.}
				\end{cases} \\
			\end{split}
		\end{equation*}
\end{itemize}

Therefore, an $\A^{a} \B^{b}$ hyperedge is created with probability $\pi_{\A}(a - 1, b)$ when a free $\A$-vertex joins an $\A^{a-1} \B^{b}$ hyperedge, and it is destroyed with probability $\pi_{\A}(a, b)$ when a free $\A$-vertex joins an $\A^{a} \B^{b}$ hyperedge.

In total, we obtain that the total expected change in $[\A^{a}\B^{b}]$ given that an $\A$-vertex is leaving a hyperedge $\A^{m}\B^{n}$ and joining another hyperedge is given as
\begin{equation}
	- \delta_{m, a}\delta_{n, b} + \delta_{m, a+1}\delta_{n, b} + \pi_{\A}(a-1, b) - \pi_{\A}(a, b).
\end{equation}
Hence, taking into account also the converse option when a $\B$-vertex is leaving the hyperegde, we obtain that
\openwidetext
\begin{align}
	\begin{split}
		\Delta [\A^{a}\B^{b}]_{(h, \text{adapt.})} & =
		\begin{aligned}[t]
			 & \mathrelphantom{+} \frac{m}{m+n}
			\left(- \delta_{m, a}\delta_{n, b} + \delta_{m, a+1}\delta_{n, b} + \pi_{\A}(a-1, b) - \pi_{\A}(a, b)\right)  \\
			 & + \frac{n}{m+n}
			\left(- \delta_{m, a}\delta_{n, b} + \delta_{m, a}\delta_{n, b+1} + \pi_{\B}(a, b-1) - \pi_{\B}(a, b) \right) \\
		\end{aligned} \\
													 & = - \delta_{m, a} \delta_{n, b}
		\begin{aligned}[t]
			 & + \frac{m}{m+n}
			\left(\delta_{m, a+1}\delta_{n, b} + \pi_{\A}(a-1, b) - \pi_{\A}(a, b)\right)   \\
			 & + \frac{n}{m+n}
			\left(\delta_{m, a}\delta_{n, b+1} + \pi_{\B}(a, b-1) - \pi_{\B}(a, b) \right). \\
		\end{aligned}
	\end{split}
\end{align}
Finally, summing this over all active hyperedges (cf.~Eq.~\eqref{eq:mean-field-equation}) we obtain for the change in the expected number of $\A^{a}\B^{b}$ hyperedges for $2 \leq a + b \leq K$ due to adaptation
\begin{equation}
	\begin{split}
		\Delta [\A^{a} \B^{b}]_{(\text{adapt.})}
		 & = \sumactiveedges{} \Delta [\A^{a} \B^{b}]_{(\A^{m} \B^{n},\text{adapt.})}                                                                                                                      \\
		 & = \frac{a+1}{a+b+1} [\A^{a + 1} \B^{b}] \indicator{\bar{Q}^K}(a+1, b) + \frac{b + 1}{a + b + 1} [\A^{a} \B^{b+1}] \indicator{\bar{Q}^K}(a, b + 1) - [\A^{a} \B^{b}] \indicator{\bar{Q}^K}(a, b) \\
		 & \mathrelphantom{=} {} + \sum_{(m, n) \in \bar{Q}^K} [\A^{m} \B^{n}] \left(\frac{m}{m + n} (\pi_{\A}(a - 1, b) - \pi_{\A}(a, b)) + \frac{n}{m + n} (\pi_{\B}(a, b - 1) - \pi_{\B}(a, b))\right), \\
		\end{split}
	\label{eq:mean-field-equation_adaptation}
\end{equation}
\closewidetext
where we used that $\sum_{(m, n) \in \bar{Q}^{K}} \delta_{m, a}\delta_{n, b} = \indicator{\bar{Q}^{K}}(a, b)$.

\subsubsection{Propagation Contribution}

For $\X \in \{\A, \B\}$ let $\eta_{\X}(m,n)$ denote the probability that an $\X$-vertex propagates its opinion in an active hyperedge $\A^{m}\B^{n}$. In particular,
\begin{itemize}[leftmargin=*]
	\item under proportional voting, $\eta_{\A}(m, n) = \frac{m}{m + n}$ and $\eta_{\B}(m, n) = \frac{n}{m + n}$ and
	\item under majority voting, $\eta_{\A}(m, n) = \Theta(m - n)$ and $\eta_{\B}(m, n) = \Theta(n - m)$, where $\Theta$ is the Heaviside step function with the convention that $\Theta(0) := \frac{1}{2}$.
\end{itemize}
Note that under both rules, $\eta_{\X} \equiv \frac{1}{2}$ on the diagonal, i.e., when $m = n$.

As before, we will assume that a propagation event takes place in an active hyperedge $h \in \{\A^{m} \B^{n}\}$ ($1 \leq m,n$). In contrast to the previous section, now we also need to consider the change to the overall number of $\A$- and $\B$-vertices. If an $\A$-vertex propagates its opinion in $h$, the number of $\A$-vertices increases by $n$, and the number of $\B$-vertices consequently decreases by the same amount. Taking into account also the converse option when a $\B$-vertex propagates its opinion, the overall expected change is given as
\begin{equation}
	\begin{split}
		\Delta [ \A ]_{(\text{prop.})} & = \sum_{(m, n) \in \bar{Q}^{K}} \hspace{-0.2cm} [\A^{m} \B^{n}] (n \, \eta_{\A}(m, n) - m \, \eta_{\B}(m,n))  \\
		\Delta [ \B ]_{(\text{prop.})} & = \sum_{(m, n) \in \bar{Q}^{K}} \hspace{-0.2cm} [\A^{m} \B^{n}] (m \, \eta_{\B}(m, n) - n \, \eta_{\A}(m,n)).
	\end{split}
	\label{eq:vertex_propagation_terms}
\end{equation}

Next, we need to compute the change to the expected number of $\A^{a} \B^{b}$ hyperedges with $2 \leq a + b \leq K$. Consider the case when an $\A$-vertex propagates its opinion. If the active hyperedge $h$ already had the form $\A^{a} \B^{b}$, i.e., if $a = m$ and $b = n$, the number of $\A^{a} \B^{b}$ hyperedges decreases by one with certainty. On the other hand, if $a = n + m$ and $b = 0$, in other words, if the $\A^{a} \B^{b}$ hyperedge was inactive and had the same number of vertices as $h$, an $\A^{a} \B^{b}$ hyperedge is created. The left part of Fig.~\ref{fig:propagation_transitions} illustrates those transitions.

In addition to those first-order terms, a propagation event in $h$ can change the number of $\A^{a} \B^{b}$ hyperedges through second-order effects, as illustrated in the right part of Fig.~\ref{fig:propagation_transitions}. For example, consider two hyperedges, $h \in \{\A \B\}$ and $h' \in \{\A \B^{2}\}$, and assume that they intersect in a $\B$-vertex to form a motif $\A (\B) \A \B$. If the $\A$-opinion propagates in the left hyperedge $h$, the motif is transformed as $\A (\B)\A \B \rightarrow \A (\A) \A \B$. This way, the right hyperedge $h'$ is converted to $\A^{2} \B$ even though this hyperedge was not directly targeted by the propagation event.

In general, such second-order effects occur whenever two hyperedges intersect in at least one vertex whose state is changed by the propagation effect. In other words, whenever an $\A$-opinion propagates in $h$, all hyperedges that intersect with $h$ in at least one $\B$-vertex are affected too, and vice versa. On hypergraphs, the situation is additionally complicated by the fact that hyperedges can intersect not just in one, but in several vertices, and several of those can change their state. Therefore, to account for all possible ways to create or destroy an $\A^{a} \B^{b}$ hyperedge, we need to sum over all possible intersection sets with other hyperedges.

To formalize this, let $\alpha$ and $\beta$ denote the sets of indices of the $\A$- and $\B$-vertices in $h$, respectively, and let $\alpha' \subseteq \alpha$ and $\beta' \subseteq \beta$ denote the indices of the vertices that lie in the intersection. This way, if an $\A$-opinion propagates in $h$, a total of $|\beta'|$ $\B$-vertices in the incident hyperedge are converted into $\A$-vertices. Therefore, if the incident hyperedge had the form $\A^{a - |\beta'|} \B^{b + |\beta'|}$, it is converted to $\A^{a} \B^{b}$ by the event. The total number of hyperedges that intersect with $h$ and have the required form can be expressed as
\begin{equation*}
	[\A^{a-|\alpha'|-|\beta'|} \B^{b} (\A_{\alpha'}^{|\alpha'|} \B_{\beta'}^{|\beta'|}) \A_{\alpha\setminus\alpha'}^{m-|\alpha'|} \B_{\beta\setminus\beta'}^{n-|\beta'|}].
\end{equation*}
On the other hand, if the incident hyperedge already had the form $\A^{a} \B^{b}$, it is destroyed by the event. Together with the first-order terms, the total change induced by a propagation of the $\A$-opinion in $h$ is equal to

\openwidefigure
\begin{figure}[H]
	\begin{minipage}[c][5cm]{\textwidth}
    \centering
    \begin{tikzpicture}[empty vertex/.style={coordinate,circle,inner sep=0,minimum size=1.5mm}]

        \begin{scope}
            \node[empty vertex, align=center, font=\footnotesize] (start) at (-1, 0) {$\A$ propagates in\\$h \in \{\A^m \B^n\}$};
            \node[empty vertex] (first_order_left) at (0, 0) {};
            \node[empty vertex] (first_order_right) at (2, 0) {};
            \node[empty vertex] (second_order_left) at (4.2, 0) {};
            \node[empty vertex] (second_order_right) at (8, 0) {};
            \node[empty vertex] (center_gray_field) at (12.0, 0) {};

            \draw[->] (first_order_left) -- node[above, fill=white, text width=1.0cm, align=center] {\scriptsize $m = a,$ $n = b$} ($(first_order_right) + (0, 1)$);
            \draw[->] (first_order_left) -- node[below, fill=white, text width=1.3cm, align=center] {\scriptsize $m + n = a,$ $0 = b$} ($(first_order_right) - (0, 1)$);

            \node[fill=black,opacity=0.1,rectangle,anchor=center,minimum width=1.1cm,minimum height=3.5cm,xshift=1cm] at (first_order_right) {};

            \node [above, align=center, font=\footnotesize] at ($(first_order_right) + (1, 1.75)$) {change in\\$[\A^a \B^b]$};
            \node at ($(first_order_right) + (1, 1)$) {$+1$};
            \node at ($(first_order_right) + (1, -1)$) {$-1$};

            \draw [decorate,decoration={calligraphic brace,amplitude=6pt},ultra thick] ($(second_order_left) + (-0.3, 1.5)$) -- ($(second_order_left) + (-0.3, -1.5)$);

            \draw[->] (second_order_left) -- node[above, fill=white, text width=2.3cm, align=center] {\scriptsize $h$ overlaps with $h'~\in~\{\A^{a - \xi} \B^{b + \xi}\}$ and $\B^\xi \in h' \cap h$} ($(second_order_right) + (0, 1)$);
            \draw[->] (second_order_left) -- node[below, fill=white, text width=2cm, align=center] {\scriptsize $h$ overlaps with $h'~\in~\{\A^{a} \B^{b}\}$ and $\B \in h' \cap h$} ($(second_order_right) - (0, 1)$);

            \node[fill=black,opacity=0.1,rectangle,anchor=center,minimum width=7.5cm,minimum height=3.5cm] at (center_gray_field) {};

            \node [above, align=center, font=\footnotesize] at ($(center_gray_field) + (0, 1.75)$) {change in\\$[\A^a \B^b]$};
            \node at ($(center_gray_field) + (0, 1)$) {$+ \displaystyle\sum_{\ldots} [\A^{a-|\alpha'|-|\beta'|} \B^{b} (\A_{\alpha'}^{|\alpha'|} \B_{\beta'}^{|\beta'|}) \A_{\alpha\setminus\alpha'}^{m-|\alpha'|} \B_{\beta\setminus\beta'}^{n-|\beta'|}]$};
            \node at ($(center_gray_field) + (0, -1)$) {$- \displaystyle\sum_{\ldots} [\A^{a - |\alpha'|} \B^{b - |\beta'|} (\A_{\alpha'}^{|\alpha'|} \B_{\beta'}^{|\beta'|}) \A_{\alpha \setminus \alpha'}^{m - |\alpha'|} \B_{\beta \setminus \beta'}^{n - |\beta'|}]$};
        \end{scope}
    \end{tikzpicture}
\end{minipage}
	\caption{All possible changes in $[\A^{a} \B^{b}]$ caused by the $\A$-opinion propagating in the hyperedge $h \in \{\A^{m} \B^{n}\}$. The corresponding diagram for a $\B$-vertex leaving can be obtained analogously.}
	\label{fig:propagation_transitions}
\end{figure}
\closewidefigure

\openwidetext
\begin{align}
	\begin{split}
		 & - \delta_{m, a} \delta_{n, b}(1 - \delta_{b, 0}) + \delta_{m + n, a} \delta_{b, 0} \\[1.5ex]
		 & + \hspace{-1cm} \sumsubsetsAplus{} \hspace{-0.7cm}
		[\A^{a - |\alpha'| - |\beta'|} \B^{b} (\A_{\alpha'}^{|\alpha'|} \B_{\beta'}^{|\beta'|}) \A_{\alpha \setminus \alpha'}^{m - |\alpha'|} \B_{\beta \setminus \beta'}^{n - |\beta'|}] -
		\hspace{-0.7cm} \sumsubsetsAminus{} \hspace{-0.7cm}
		[\A^{a - |\alpha'|} \B^{b - |\beta'|} (\A_{\alpha'}^{|\alpha'|} \B_{\beta'}^{|\beta'|}) \A_{\alpha \setminus \alpha'}^{m - |\alpha'|} \B_{\beta \setminus \beta'}^{n - |\beta'|}].
	\end{split}
\end{align}
\closewidetext
Here, the conditions in the sums prevent against non-physical edge cases. First, we need to ensure that $\beta'$ is not an empty set, because if the intersection set does not contain any $\B$ vertices, then not a single vertex in the incident hyperedge will change. Next, we need to exclude an exact overlap between the two hyperedges since we do not allow multi-edges. Hence, we require $|\alpha'| + |\beta'| < \max(a + b, m + n)$. Finally, the condition in the third line of the sum ensures that the number of vertices does not become negative.

Taking again into account also the converse option that a $\B$-vertex propagates its opinion, we obtain that
\openwidetext

\begin{align}
	\begin{split}
		\Delta [\A^{a} \B^{b}]_{(h,~\text{prop.})} =
		 & - \delta_{m, a} \delta_{n, b}(1 - \delta_{b, 0}) + \delta_{m + n, a} \delta_{b, 0}     \\[1.5ex]
		 & + \eta_{\A}(m, n)
		\begin{aligned}[t]
			\left( \mathrelphantom{-} \sumsubsetsAplus{} \right.
			 & [\A^{a - |\alpha'| - |\beta'|} \B^{b} (\A_{\alpha'}^{|\alpha'|} \B_{\beta'}^{|\beta'|}) \A_{\alpha \setminus \alpha'}^{m - |\alpha'|} \B_{\beta \setminus \beta'}^{n - |\beta'|}] \,                                                                      \\
			- \sumsubsetsAminus{}
			 & [\A^{a - |\alpha'|} \B^{b - |\beta'|} (\A_{\alpha'}^{|\alpha'|} \B_{\beta'}^{|\beta'|}) \A_{\alpha \setminus \alpha'}^{m - |\alpha'|} \B_{\beta \setminus \beta'}^{n - |\beta'|}] \left. \vphantom{\sumsubsetsAminus{}} \right)
		\end{aligned} \\
		 & + \eta_{\B}(m, n)
		\begin{aligned}[t]
			\left( \mathrelphantom{-} \sumsubsetsBplus{} \right.
			 & [\A^{a} \B^{b - |\alpha'| - |\beta'|} (\A_{\alpha'}^{|\alpha'|} \B_{\beta'}^{|\beta'|}) \A_{\alpha \setminus \alpha'}^{|\alpha \setminus \alpha'|} \B_{\beta \setminus \beta'}^{|\beta \setminus \beta'|}] \,                                             \\
			- \sumsubsetsBminus{}
			 & [\A^{a - |\alpha'|} \B^{b - |\beta'|} (\A_{\alpha'}^{|\alpha'|} \B_{\beta'}^{|\beta'|}) \A_{\alpha \setminus \alpha'}^{|\alpha \setminus \alpha'|} \B_{\beta \setminus \beta'}^{|\beta \setminus \beta'|}] \left. \vphantom{\sumsubsetsBminus{}} \right).
		\end{aligned}
	\end{split}
\end{align}

Summing this again over all active hyperedges (cf. Eq.~\eqref{eq:mean-field-equation}), we obtain for change in the expected number of $\A^{a}\B^{b}$ hyperedges for $2 \leq a + b \leq K$ due to propagation,
%
\begin{equation}
	\begin{split}
		\Delta [\A^{a} \B^{b}]_{(\text{prop.})} & = \sumactiveedges{} \Delta [\A^{a} \B^{b}]_{(h,~\text{prop.})}              \\
											    & =
		\begin{aligned}[t]
			 & - [\A^{a} \B^{b}] \indicator{\bar{Q}^K}(a, b)
			+ \sum_{1 \leq \mu \leq a-1} \eta_{\A}(a - \mu, \mu) [\A^{a - \mu} \B^{\mu}] \delta_{b, 0}
			+ \sum_{1 \leq \nu \leq b-1} \eta_{\B}(\nu, b - \nu) [\A^{\nu} \B^{b - \nu}] \delta_{a, 0}                        \\[1.5ex]
			 & + \sum_{(m, n) \in \bar{Q}^K} \eta_{\A}(m,n)
			\begin{aligned}[t]
				\left( \mathrelphantom{-} \sumintersectionsAplus \right. & (1 + \delta_{m, a - \nu}\delta_{n - \nu, b})
				[\A^{a - \mu - \nu} \B^{b} (\A^{\mu} \B^{\nu}) \A^{m-\mu} \B^{n-\nu}]                                         \\
				- \sumintersectionsAminus                                & (1 + \delta_{m, a}\delta_{n, b})
				[\A^{a - \mu} \B^{b - \nu} (\A^{\mu} \B^{\nu}) \A^{m-\mu} \B^{n-\nu}]
				\left. \vphantom{\sumintersectionsAplus} \right)
			\end{aligned} \\
			 & + \sum_{(m,n) \in \bar{Q}^K} \eta_{\B}(m, n)
			\begin{aligned}[t]
				\left( \mathrelphantom{-} \sumintersectionsBplus \right. & (1 + \delta_{m - \mu, a} \delta_{n, b - \mu})
				[\A^{a} \B^{b-\mu-\nu} (\A^{\mu} \B^{\nu}) \A^{m - \mu} \B^{n - \nu}]                                          \\
				- \sumintersectionsBminus                                & (1 + \delta_{m, a} \delta_{n, b})
				[\A^{a-\mu} \B^{b-\nu} (\A^{\mu} \B^{\nu}) \A^{m-\mu} \B^{n-\nu}]
				\left. \vphantom{\sumintersectionsBplus} \right),
			\end{aligned}                                                                                                      \\
		\end{aligned}
	\end{split}
	\label{eq:mean-field-equation_propagation}
\end{equation}
\closewidetext
where combinatorial coefficients like $1 + \delta_{m, a - \nu}\delta_{n, b + \nu}$ account for the double-counting of symmetric motifs.

\subsubsection{The Complete Equations}

This completes the derivation of the mean-field description. The final equations can be obtained by inserting the adaptation and propagation terms from Eq.~\eqref{eq:mean-field-equation_adaptation} and \eqref{eq:mean-field-equation_propagation} into Eq.~\eqref{eq:mean-field-equation}. We include those for convenience in the \href{sec:appendix}{Appendix}.

\paragraph{Reduction to the Special Case of a Graph} The special case of a graph can be obtained by setting $K = 2$. In this case, the index set of active hyperedges is simply equal to $\bar{Q}^2 = \{(1, 1)\}$. The joining probabilities $\pi_{\A}$ and $\pi_{\B}$ reduce to
\begin{itemize}[leftmargin=*]
	\item \parbox{\linewidth}{\begin{align*}
			\pi_{\A}(1, 0) & = \pi_{\B}(1, 0) = \frac{[\A]}{N}, \\
			\pi_{\A}(0, 1) & = \pi_{\B}(0, 1) = \frac{[\B]}{N}
		\end{align*}}
		under the rewire-to-random rule and
	\item \parbox{\linewidth}{\begin{align*}
			\pi_{\A}(1, 0) & = \pi_{\B}(0, 1) = 1, \\
			\pi_{\A}(0,1)  & = \pi_{\B}(1, 0) = 0
		\end{align*}}
		under the rewire-to-same rule.
\end{itemize}

Altogether, we obtain for the graph model
\begin{equation}
	\begin{split}
		\derivative{t} [\A]   & = 0                                                                        \\[1.5ex]
		\derivative{t} [\B]   & = 0                                                                        \\[1.5ex]
		\derivative{t} [\A\A] & =
		\begin{aligned}[t]
			 & \frac{1 - p (1 - \pi_{\A}(1, 0))}{2} [\A \B]                    \\
			 & + \frac{1 - p}{2} \, \bigl(2 [\A (\B) \A] - [\A (\A) \B] \bigr) \\
		\end{aligned} \\[1.5ex]
		\derivative{t} [\A\B] & =
		\begin{aligned}[t]
			 & - \left(1 - \frac{p (\pi_{\A}(0, 1) + \pi_{\B}(1, 0))}{2} \right) [\A \B] \\
			 & + \frac{1 - p}{2}
			\begin{aligned}[t]
			\bigl( & [\B (\B) \A] - 2 [\A (\B) \A]        \\
			+      & [\A (\A) \B] - 2 [\B (\A) \B] \bigr)
			\end{aligned}
		\end{aligned}                                              \\[1.5ex]
		\derivative{t} [\B\B] & =
		\begin{aligned}[t]
			 & \frac{1 - p (1 - \pi_{\B}(0, 1))}{2} [\A \B]                     \\
			 & + \frac{1 - p}{2} \, \bigl(2 [\B (\A) \B] - [\B (\B) \A] \bigr). \\
		\end{aligned}
	\end{split}
\end{equation}

The fact that these equations do not explicitly depend on $\eta_{\A}$ and $\eta_{\B}$ anymore shows that the majority and proportional voting rules are equivalent on graphs. Importantly, one confirms that these equations are the same as the ones that have been derived elsewhere~\parencite{durrett2012graph,demirel2014momentclosure}.

\subsubsection{Moment-Closure Approximation}

The mean-field equations we have derived above are only the first in a hierarchy of equations for ever larger network moments. In order to obtain a closed system that is amenable to further analysis and can also be solved numerically, one generally utilizes moment-closure relations with which higher-order moments are expressed through lower-order ones~\parencite{pellis2015exact,kuehn2016moment,wuyts2022meanfield}.

For the mean-field equations here, we propose
\begin{equation}
	\begin{split}
		 & \mathrelphantom{=} [\A^{a-\mu} \B^{b-\nu} (\A^{\mu} \B^{\nu}) \A^{m-\mu} \B^{n-\nu}]                                                                                                \\
		 & \approx \frac{\binom{a}{\mu} \binom{b}{\nu} \binom{m}{\mu} \binom{n}{\nu}}{(1 + \delta_{a, m} \delta_{b, n})} \frac{[\A^{a} \B^{b}][\A^{m} \B^{n}]}{\binom{[\A]}{\mu}\binom{[\B]}{\nu}} \\
	\end{split}
	\label{eq:closure}
\end{equation}
as closure relation, which is a generalization of the known pair-approximation on graphs~\parencite{rand1999correlation,sharkey2006pairlevel,demirel2014momentclosure}.

The closure can be obtained by the following arguments. We want to estimate the number of $\A^{a} \B^{b}$ and $\A^{m} \B^{n}$ hyperedges which intersect exactly in $\mu$ $\A$-vertices and $\nu$ $\B$-vertices. Start counting from the left $\A^{a} \B^{b}$ hyperedge. The number of such hyperedges is equal to $[\A^{a} \B^{b}]$. Furthermore, there are $\binom{a}{\mu} \binom{b}{\nu}$ ways to choose the set of intersection vertices from each $\A^{a} \B^{b}$ hyperedge. Fix one of those sets. Let $\tanglebr{q_{\A^{\mu} \B^{\nu}}}$ denote the average number of additional hyperedges incident on this set of vertices, similarly to the mean excess degree $\tanglebr{q}$ on graphs. To obtain the closure, we need to estimate how many of those additional incident hyperedges have the form $\A^{m} \B^{n}$.

At this point, we need to make an assumption typical for pair-approximation closures that hyperedges of all types are distributed uniformly in the network. In other words, we will assume that a neighboring hyperedge has the same probability to be of a certain type as a random hyperedge incident on a set of random $\mu$ $\A$-vertices and $\nu$ $\B$-vertices chosen uniformly at random from the whole hypergraph. Intuitively, by making this approximation, we discard the information that we chose the set of intersection vertices because they all belong to the same hyperedge $\A^{a} \B^{b}$. Denote the average number of hyperedges incident on a set of randomly chosen $\mu$ $\A$-vertices and $\nu$ $\B$-vertices with $\tanglebr{k_{\A^{\mu} \B^{\nu}}}$. This quantity generalizes the mean degree $\tanglebr{k}$ on graphs. We want to calculate the probability that a random hyperedge incident on the set of vertices has the form $\A^{m} \B^{n}$. The total number of ways to choose a set of vertices from the hypergraph and one hyperedge incident on the set is equal to $\binom{[\A]}{\mu} \binom{[\B]}{\nu} \tanglebr{k_{\A^{\mu} \B^{\nu}}}$. Of those, $\binom{m}{\mu} \binom{n}{\nu} [\A^{m} \B^{n}]$ hyperedges have the required form $\A^{m} \B^{n}$. After collecting all factors, we obtain
\begin{equation}
	\frac{\binom{a}{\mu} \binom{b}{\nu} \binom{m}{\mu} \binom{n}{\nu}}{1 + \delta_{a, m} \delta_{b, n}}
	\frac{[\A^{a} \B^{b}] [\A^{m} \B^{n}] \tanglebr{q_{\A^{\mu} \B^{\nu}}}}{\binom{[\A]}{\mu}\binom{[\B]}{\nu} \tanglebr{k_{\A^{\mu} \B^\nu}}},
\end{equation}
where the term $1 + \delta_{a, m} \delta_{b, n}$ is needed to account for double-counting of symmetric motifs.

In adaptive models on graphs, the approximation $\tanglebr{q} \approx \tanglebr{k}$ gives good empirical results and is exact on non-adaptive Erd\H{o}s-R\'enyi graphs~\parencite{gross2009adaptive, kimura2008coevolutionary, vazquez2008generic}. In the same spirit, we will approximate $\tanglebr{q_{\A^{\mu}\B^{\nu}}} \approx \tanglebr{k_{\A^{\mu}\B^{\nu}}}$ to obtain the final closure in Eq.~\eqref{eq:closure}.

\subsubsection{Evolution of the Magnetization}

An important observable of the dynamics is the magnetization of the system, i.e., the difference between the density of opinions, defined as
\begin{equation}
	\mathfrak{m} := \frac{1}{N} \left([\A] - [\B]\right),
\end{equation}
such that the magnetization is positive if opinion $\A$ dominates and negative if opinion $\B$ dominates.

For its dynamics, we have that (cf. Eq.~\eqref{eq:mean-field-equation})
\begin{equation}
	\begin{split}
	\derivative{t} \mathfrak{m} & = \frac{1-p}{N} (\Delta [ \A ]_{(\text{prop.})} - \Delta [ \B ]_{(\text{prop.})})                                                                                 \\
								& = \frac{2(1-p)}{N} \hspace{-0.2cm} \sum_{(m, n) \in \bar{Q}^{K}} \hspace{-0.2cm} [\A^{m} \B^{n}] (n \, \eta_{\A}(m, n) - m \, \eta_{\B}(m,n)).
	\end{split}
\end{equation}

Under proportional voting, when $\eta_{\A}(m, n) = \frac{m}{m + n}$ and $\eta_{\B}(m, n) = \frac{n}{m + n}$, both terms in the sum cancel each other out so that the magnetization is conserved. However, this is not the case in general and specifically under majority voting, when $\eta_{\A}(m, n) = \Theta(m - n)$ and $\eta_{\B}(m, n) = \Theta(n - m)$. In this case, we obtain that
\begin{equation}
	\begin{split}
		\derivative{t} \mathfrak{m} & = \frac{2(1-p)}{N} \hspace{-0.2cm} \sum_{\substack{(m, n) \in \bar{Q}^K \\ m > n}} n \left( [\A^{m} \B^{n}] - [\A^{n} \B^{m}] \right).
	\end{split}
	\label{eq:magnetization_stability}
\end{equation}

If the magnetization is equal to zero, then $\A$- and $\B$-vertices should behave identically since all update rules are symmetric with respect to the inversion of all states. In particular, the number of hyperedges of each kind should not change with the inversion of all states in expectation, $[\A^{m} \B^{n}] = [\A^{n} \B^{m}]$. If we perturb the magnetization in the direction of $\A$ away from $0$, one should expect that this equilibrium shifts in favor of hyperedges containing more $\A$-vertices than $\B$-vertices, i.e., $[\A^{m} \B^{n}] > [\A^{n} \B^{m}]$ if $m > n$. Therefore, the whole right-hand side in Eq.~\eqref{eq:magnetization_stability} is positive and thus $\derivative{t} \mathfrak{m} > 0$. Similarly, if we perturb the magnetization in the direction of $\B$, one would observe the opposite and thus $\derivative{t} \mathfrak{m} < 0$. Therefore, zero magnetization is unstable under majority voting.

\subsection{Numerical Investigations}

In the following section, we will further our investigations into the dynamics of the adaptive voter model on a hypergraph by comparing simulations (see \href{sec:appendix}{Appendix} for implementation details) to a numerical integration of the analytical mean-field description.

\openwidefigure
\begin{figure}[H]
	\centering
	\includegraphics{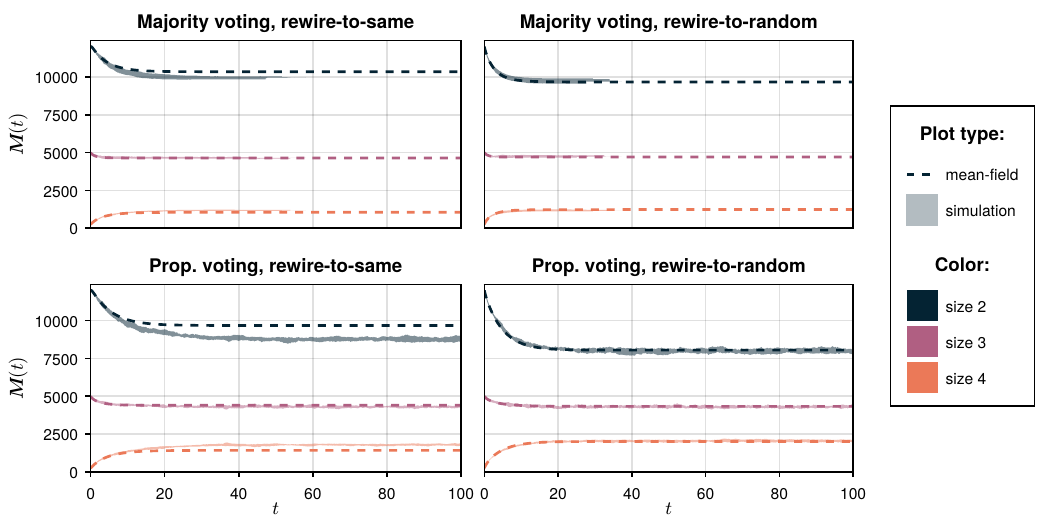}
	\caption{Total number of hyperedges of every size for all four combinations of rules. Area plots: simulated results from 12 realizations, dashed lines: analytical results. The distribution of hyperedges quickly reaches a stable state that is remarkably stable against stochastic fluctuations. Under the majority voting rule, the simulation depletes all active hyperedges at $t \approx 30$ for the rewire-to-same rule and at $t \approx 50$ for rewire-to-random. Parameters: $N = \num{10000}$, $\mathbf{M}(0) = (\num{12000}, \num{5000}, \num{250})$, $p = 0.5$, $\tanglebr{\mathfrak{m}_{0}} = 0.5$.}
	\label{fig:hyperedges_all_four}
\end{figure}
\closewidefigure

Unless mentioned otherwise, all analyses are performed on a hypergraph with $N = \num{10000}$ vertices and the maximum size of hyperedges fixed at $K = 4$. The initial distribution of hyperedges is set to $\mathbf{M}(0) = (M_{2}, M_{3}, M_{4}) = (\num{12000}, \num{5000}, \num{250})$. Here, we follow the assumption made in \cite{papanikolaou2022consensus} that groups of greater size are less frequent because they are more costly for the individuals to form and maintain. Additionally, from a practical perspective, large hyperedges are computationally expensive to simulate. The exact numbers are chosen in such a way as to give an average degree of ${\tanglebr{k} = \frac{1}{N}\sum_{i=2}^K M_{i} \cdot i = 4}$, which is a common choice in the literature on graphs \cite{vazquez2008generic, durrett2012graph, zschaler2012early}.

Since the rules do not conserve the number of hyperedges, it is interesting to first check how $\mathbf{M}$ changes over time. Fig.~\ref{fig:hyperedges_all_four} shows the evolution of the total number of hyperedges of every size for all four combinations of rules: majority voting or proportional voting and rewire-to-same or rewire-to-random. All four cases start from an initial magnetization $\tanglebr{\mathfrak{m}_{0}} = 0.5$ that favors the $\A$-opinion. The rewiring probability is set to $p = 0.5$.

In all four cases, the distribution of hyperedges quickly converges to a stable state. This distribution is very stable against random fluctuations: Fig.~\ref{fig:hyperedges_all_four} shows trajectories from 12 different realizations, but the deviations between them are minor. The most noticeable difference between the four plots is that the simulation results end abruptly in both majority voting panels at $t < 60$ but not in proportional voting panels. The simulation terminated early in the case of majority voting because the system depleted all active links and converged to an absorbing state. On the other hand, under proportional voting, the simulation continues to run for a long time with the distribution of hyperedges remaining constant.

The analytical solution matches well the behavior of the simulation, with the largest deviation observed in the case of proportional voting and rewire-to-same rule.

\begin{figure}[H]
	\centering
	\includegraphics{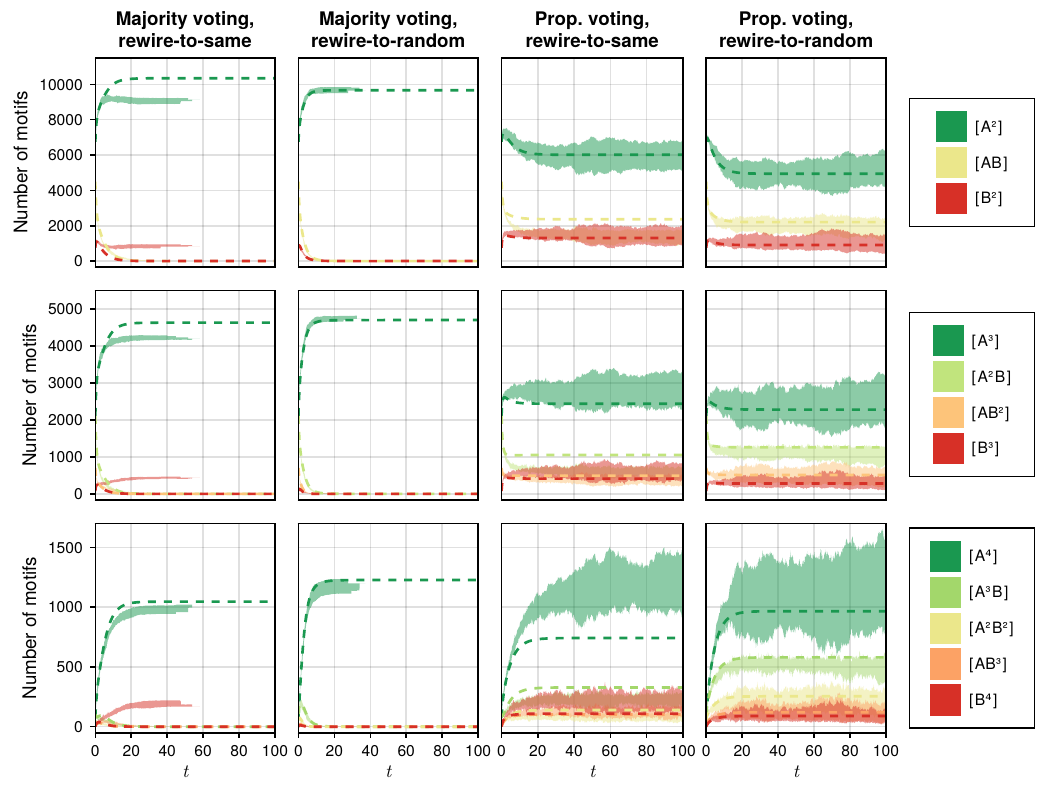}
	\caption{Analytical solution for the evolution of every moment compared to simulated trajectories. Motifs of the same size are displayed in the same row. Note that the scale of the y-axis changes between rows. Area plots: simulated results from 12 realizations, dashed lines: analytical results. Under proportional voting, the number of individual motifs exhibits much higher variance than the total number of hyperedges; under majority voting, the variance remains low. Parameters: $N = \num{10000}$, $\mathbf{M}(0) = (\num{12000}, \num{5000}, \num{250})$, $p = 0.5$, $\tanglebr{\mathfrak{m}_{0}} = 0.5$.}
	\label{fig:motif_plot}
\end{figure}

On a finer level, the number of hyperedges of a particular size can be split into the number of individual motifs of this size. Fig.~\ref{fig:motif_plot} shows the time evolution of the motifs from the same simulation as in Fig.~\ref{fig:hyperedges_all_four} and the corresponding analytical solutions. For convenience, the motifs are grouped into different panels by their size. This way, the sum of all motifs in a panel is equal to the total number of hyperedges of the corresponding size; for example, the sum of all motifs in the top row corresponds to the black line in Fig.~\ref{fig:hyperedges_all_four}.

Overall, the numerical evolution of motifs undergoes much higher variance than the evolution of the total number of hyperedges, especially in the case of proportional voting. The analytical solution captures the qualitative behavior of the simulation, but deviates from the numerical values for some motifs (see, for example, the size four motifs in proportional voting and rewire-to-same).

\openwidefigure
\closewidefigure

\begin{figure}[H]
	\centering
	\includegraphics{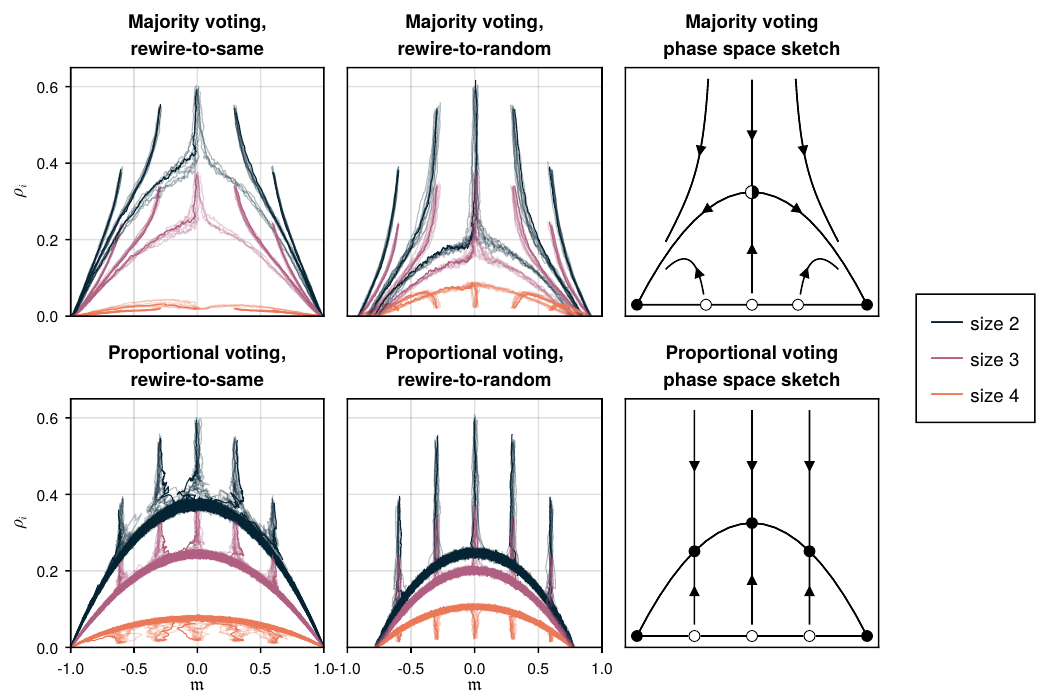}
	\caption{The slow manifold plot for all four combinations of rules. Initial conditions start from different magnetization values $\tanglebr{\mathfrak{m}_0} \in \{-0.6, -0.3, 0, 0.3, 0.6\}$; 12 simulations were performed for every initial magnetization. Under majority voting, the initial bias in magnetization is amplified, meaning that the trajectories converge rapidly to the absorbing roots of the parabola. Under proportional voting, the trajectories instead slowly diffuse on the parabola. The panels in the third column illustrate this behavior qualitatively. Parameters: $N = \num{10000}$, $\mathbf{M}(0) = (\num{12000}, \num{5000}, \num{250})$. $p = 0.1$ for majority voting and $p = 0.7$ for proportional voting.}
	\label{fig:slow_manifold}
\end{figure}

To understand the difference in run time between majority and proportional voting, it is helpful to plot the density of active hyperedges against the magnetization. It is known from graph models that the trajectories form a parabola-shaped manifold when plotted this way~\parencite{kimura2008coevolutionary, vazquez2008generic, durrett2012graph}. Since there are multiple kinds of active hyperedges on hypergraphs, we will lump all active hyperedges of the same size $i$ together,
\begin{equation}
	\rho_{i} := \frac{1}{N} \sum_{m = 1}^{i-1} [\A^{m} \B^{i-m}].
\end{equation}
The result is shown in Fig.~\ref{fig:slow_manifold}. For each rule, we start the simulation at different initial magnetization values, $\tanglebr{\mathfrak{m}_{0}} \in \{-0.6, -0.3, 0, 0.3, 0.6\}$, and consider trajectories from 12 different realizations for every combination of parameters. One trajectory from every batch is highlighted in a darker color. Note that the rewiring probability is set to $p = 0.1$ under majority voting and to $p = 0.7$ under proportional voting. The reason for choosing different values of $p$ will become evident once we plot the phase transition diagram. To briefly foreshadow the results, the critical value $p_{c}$ is low under majority voting but is very close to $1$ under proportional voting, so we chose values of $p$ which best illustrate the typical shape of the parabola for $p < p_{c}$.

Let us focus on majority voting simulations first. A parabola-shaped curve is apparent, but the trajectories behave qualitatively differently compared to the graph models. Instead of falling vertically in the initial phase, the trajectories start drifting to the left or the right even before they hit the parabola. This confirms our analytical result that the initial bias in the distribution of states is amplified: If a trajectory starts from a negative magnetization, it is pushed even further toward negative magnetization, and vice versa. A sketch of the flow in the phase space in the third column illustrates this behavior.

\openwidefigure
\closewidefigure

In contrast, under proportional voting, magnetization is conserved in the mean-field and only changes due to stochastic fluctuations. Therefore, the plot looks similar to the results observed on graphs: First, the trajectories fall down until they hit the parabola, then, they slowly diffuse on the parabola until they eventually hit one of the roots.

When comparing both rewiring rules, we can observe that the roots of both rewire-to-same parabolas are fixed at $-1$ and $1$, and the roots of the rewire-to-random parabolas are located between $-1$ and $1$. This behavior exactly matches the results on graphs~\parencite{durrett2012graph}. Furthermore, the roots lie at the same points for all sizes of hyperedges, and only the height of the parabola changes. Therefore, the ratio between the number of active hyperedges of different sizes remains approximately constant at all times.

The parabola plot can be used to analyze the fragmenting phase transition. From the results on graphs, we expect the parabola to decrease with increasing $p$ and disappear entirely at the critical point $p_{c}$. Therefore, we can use the roots of the parabola, or, in other words, the final magnetization of the system in the absorbing state, as an order parameter of the system.

\begin{figure}[H]
	\centering
	\includegraphics{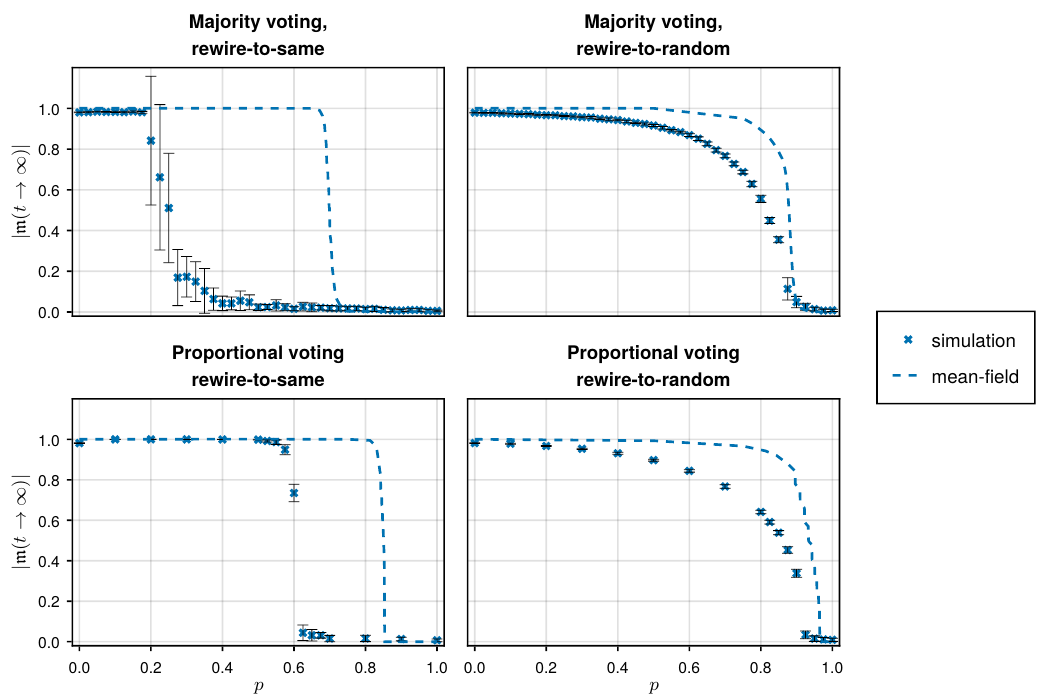}
	\caption{Absolute value of magnetization in the absorbing state of the simulation for different values of $p$. Under the rewire-to-same rule, magnetization switches from $|\mathfrak{m}| = 1$ to $|\mathfrak{m}| = 0$, indicating a first-order phase transition. In contrast, under the rewire-to-random rule, the transition is gradual, which implies a second-order phase transition. Parameters: $N = \num{10000}$, $\mathbf{M}(0) = (\num{12000}, \num{5000}, \num{250})$, $\tanglebr{\mathfrak{m}_0} = 0$.}
	\label{fig:phase_transition}
\end{figure}

To measure the final magnetization numerically, we start the simulation with an initial magnetization of $\tanglebr{\mathfrak{m}_{0}} = 0$ and let the system evolve until it reaches an absorbing state. The absolute value of the magnetization in this state is measured and averaged over 12 realizations of the simulation for every value of $p$. We follow a similar procedure to get the mean-field results. Under majority voting, the magnetization in the absorbing state can be obtained by simply solving the equations numerically until they converge to one of the roots. To break the symmetry between opinions, we start from a slightly positive initial magnetization $\mathfrak{m}_{0} = \num{0.0002}$. On the other hand, under proportional voting, we cannot obtain the magnetization directly because the mean-field trajectories never reach the roots and instead converge to the stable points on the parabola. Instead, we compute the stable states of the system for different values of $p$ to sample multiple points on the slow manifold and fit a centered parabola to all points that are greater than zero.
The result is shown in Fig.~\ref{fig:phase_transition}. Both rewire-to-same models undergo a sharp transition under variation of $p$ in which the absolute magnetization switches from $|\mathfrak{m}| = 1$ to $|\mathfrak{m}| = 0$. Therefore, the roots of the parabola are fixed at $\pm 1$ for $p < p_{c}$ under the rewire-to-same rule. On the other hand, under rewire-to-random, we observe a second-order phase transition with the absolute magnetization slowly decreasing with \parfillskip=0pt 

\openwidefigure
\closewidefigure

increasing $p$. Both results replicate the known behavior on graphs under the rewire-to-random and rewire-to-same rules~\parencite{durrett2012graph}. Also, note that near the phase transition, stochastic fluctuations are larger, which can be explained by the critical slowing down effect of the deterministic dynamics as commonly exploited in the theory of early-warning signs~\cite{wiesenfeld1985noisy,scheffer2009early,kuehn2013mathematical}.

Furthermore, we observe that the mean-field results overestimate the position of the critical point, significantly so in both rewire-to-same cases. The poor performance of the pair-approximation closure close to the critical point is a well-known problem on graphs. It is discussed in detail in~\parencite{demirel2014momentclosure}, where the authors have shown that the moment closure breaks down on the moments $[\A(\B)\A]$ and $[\B(\A)\B]$ as one approaches the fragmenting phase transition. At this point, the graph becomes highly clustered, such that the assumption of uncorrelated edges underlying the moment closure no longer holds. In contrast, moments of the type $[\B(\B)\A]$ are well-approximated by the closure for all values of $p$. To investigate whether this correlation can also explain the poor performance on hypergraphs, we performed a similar analysis done in \parencite{demirel2014momentclosure} and compared the simulated number of second-order motifs to the number of those motifs predicted by the closure. To reduce the amount of information, we focus on motifs of the form $\A^{a} (\B) \A^{n}$, $\B^{b} (\B) \A^{n}$, and $\A^{a} \B^{b} (\B) \A^{n} \B^{m}$. Furthermore, we restrict ourselves to motifs with a single vertex in the intersection as those with larger intersections are significantly less frequent.

\begin{figure}[H]
	\centering
	\includegraphics{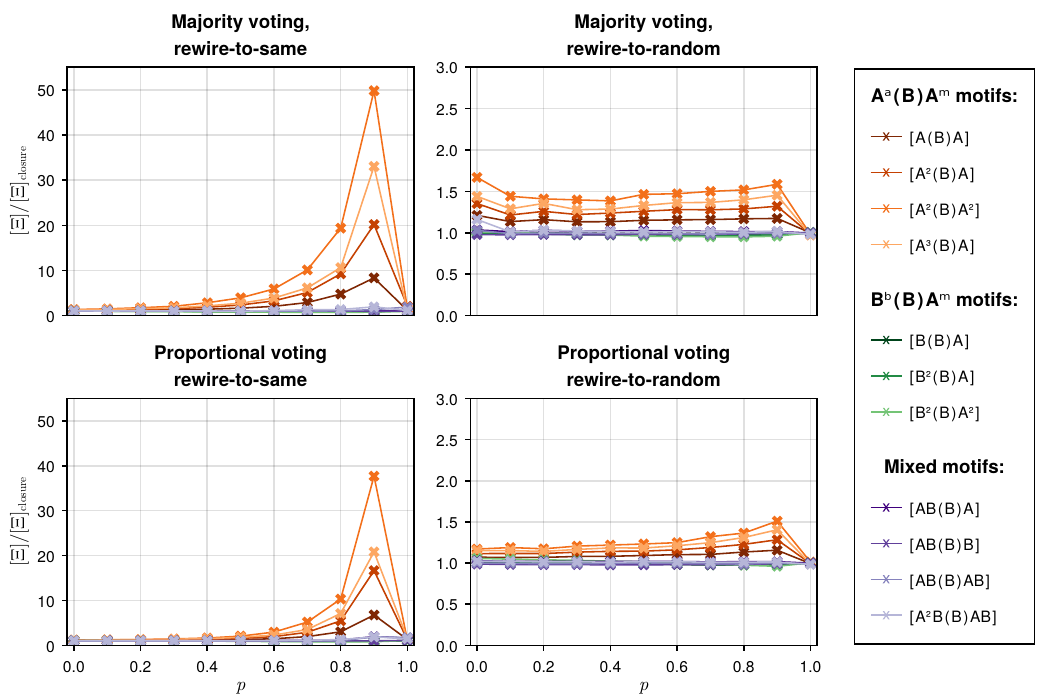}
	\caption{Performance of the moment-closure for selected second-order motifs. Note the different scales of the y-axis. Shown is the ratio between the simulated number of second-order motifs and the estimate calculated using Eq.~\eqref{eq:closure} based on the simulated number of first-order motifs. Motifs of the type $\A^{a} (\B) \A^{m}$ are poorly approximated by the closure for large values of $p$ under the rewire-to-same rule. Parameters: $N = \num{10000}$, $\mathbf{M}(0) = (\num{12000}, \num{5000}, \num{250})$, $\tanglebr{\mathfrak{m}_0} = 0.5$.}
	\label{fig:moment_closure_performance}
\end{figure}

Fig.~\ref{fig:moment_closure_performance} demonstrates the performance of the moment closure for all four combinations of rules. To compute the ratio, we evolve the system until $t_\text{end} = 10$ and record the number of all motifs in the hypergraph over time. From this data, we estimate the number of second-order moments from the first-order ones using the closure in Eq.~\eqref{eq:closure} at all points in time. Then, we take the ratio between the true and predicted values; those values are then averaged over time and over 12 different realizations.

The results in Fig.~\ref{fig:moment_closure_performance} strikingly illustrate the differences in performance between the rewire-to-same and rewire-to-random rules. Under the rewire-to-same rule, the $\A^{a} (\B) \A^{n}$-type motifs are not well approximated by the closure for large values of $p$, but this effect is much less pronounced under the rewire-to-random rule (note the different scales of the y-axes).

\openwidefigure
\closewidefigure

\section{Discussion}

In this work, we generalized the adaptive voter model to hypergraphs by introducing four different update rules: majority and proportional voting rules that govern the spread of opinions and rewire-to-same and rewire-to-random rules that control the adaptation. Furthermore, we derived mean-field equations in terms of hypergraph motifs and a corresponding moment closure on the pair level. The equations can be easily adapted to describe other similar models by modifying the joining probability $\pi_\X(a, b)$ and the propagation probability $\eta_{\X}(a, b)$.

We found that even though the adaptation rules do not conserve the number of hyperedges, the topology of the hypergraph rapidly converges toward a stable distribution of hyperedges, which is remarkably robust against stochastic fluctuations.

Furthermore, we observed a drastic difference in convergence times between majority and proportional voting. Under majority voting, the existing bias in the distribution of opinions is amplified, which forces the system to converge rapidly to an absorbing state. This finding agrees with previous work on the voter model on graphs of higher order~\parencite{horstmeyer2020adaptive, papanikolaou2022consensus}. However, under proportional voting, the dynamics exhibits similar behavior to the models on graphs~\parencite{kimura2008coevolutionary, vazquez2008generic, durrett2012graph}, in that the density of active hyperedges first converges to a parabola-shaped manifold, then slowly diffuses on it. In this metastable state, active hyperedges are not depleted and persist for a long time. The ratio between active hyperedges of different sizes stays approximately constant under our combination of rules; this stands in contrast with the model in~\parencite{horstmeyer2020adaptive} in which the system eventually runs out of triangles that can be promoted to 2-simplices.

We also investigated the fragmentation transition for all four combination of rules, with the absolute magnetization in the absorbing state as an order parameter. Similarly to the finding on graphs~\parencite{durrett2012graph}, our numerical experiments suggest a first-order discontinuous phase transition under the rewire-to-same rule and a second-order continuous phase transition under the rewire-to-random rule.

The mean-field equations accurately capture the evolution of the expected number of motifs and correctly predict the existence of a fragmenting phase transition. However, there is a significant deviation between the simulated and the mean-field critical point $p_{c}$ under the rewire-to-same rule. As Fig.~\ref{fig:moment_closure_performance} illustrates, this poor performance is caused by $\A^{a} (\B) \A^{m}$-type motifs (and, by symmetry, $\B^{b} (\A) \B^{n}$), similarly to the situation on graphs \parencite{demirel2014momentclosure}. A possible direction for further research would be to search for a better closure close to the critical point $p_{c}$ as was done on graphs~\parencite{boehme2011analytical}.

On the other hand, the closure yields accurate results under the rewire-to-random rule, with the mean-field equations accurately predicting the critical point $p_{c}$. This effect also manifests itself when looking at the evolution of specific motifs (Fig.~\ref{fig:motif_plot}), where the mean-field equations give better results under the rewire-to-random rule compared to the rewire-to-same rule.
One can speculate that rewiring vertices to random hyperedges reduces the correlations between incident hyperedges and thereby improves the performance of the closure. However, this might not hold for a different choice of parameters, for example, a different initial distribution of hyperedges $\mathbf{M}(0)$ or non-zero initial magnetization.

There are several possible directions for future research. The hypergraph setting provides a fertile ground for exploring alternative updating rules. In the case of adaptation, one could explore what happens if vertices preferentially join hyperedges where their opinion is in the majority instead of selecting a hyperedge uniformly. Similarly, the probability to select the adaptive vertex can depend on the degree of the vertex or the number of adjacent active hyperedges. In the case of propagation, proportional voting could also be implemented in such a way that vertices update their opinions independently with the same probability. Further research is needed to determine whether those variations will qualitatively change the results.

Although we derived a system of ODEs for the model, we had to solve the equations numerically because of the large number of variables involved. A computer algebra system might help to deal with the increasing complexity of the equations and provide analytical expressions for the critical point or the slow manifold. In particular, a computer algebra system might help to rigorously carry out a center manifold reduction for the moment system to study the bifurcations/phase transitions analytically. These local bifurcation results could be complemented by a detailed numerical continuation study in multiple parameters.

From an applied perspective, the predictions of the hypergraph model should be compared with real-world data before any interpretations can be drawn. In addition to that, research on the voter model provides many sources of inspiration for how to extend the hypergraph model to include even more effects observed in reality. Different options include considering noisy models~\parencite{granovsky1995noisy,carro2016noisy}, allowing more than two opinions \parencite{starnini2012ordering, pickering2016solution}, adding stubborn zealots who never change their opinion~\parencite{klamser2017zealotry}, or letting an individual change its opinion only after repeated reinforcement~\parencite{castellano2009nonlinear,redner2019reality}.

\subsection*{Data Availability Statement}
The data that support the findings of this study are openly available in figshare at \textsc{doi}:~\href{https://doi.org/10.6084/m9.figshare.c.6758496}{\texttt{10.6084/m9.figshare.c.6758496}}.

\subsection*{Acknowledgements}
The authors acknowledge the Leibniz Supercomputing Centre for providing computing time on its Linux-Cluster. A.G. acknowledges funding from the Max-Planck-Gesellschaft and the German Federal Ministry for Education and Research for the infoXpand project (031L0300A). J.M. and C.K. acknowledge funding from the Deutsche Forschungsgemeinschaft (DFG, German Research Foundation) as well as partial support from the VolkswagenStiftung via a Lichtenberg Professorship awarded to C.K.

\subsection*{Conflict of Interest}
The authors declare no conflict of interest.

\subsection*{Author Contributions}
Conceptualization: A.G., J.M., and C.K.; Formal analysis: A.G. and J.M.; Funding acquisition: C.K.; Investigation: A.G. and J.M.; Methodology: A.G., J.M., and C.K.; Software: A.G.; Supervision: J.M. and C.K.; Writing -- original draft: A.G. and J.M.; Writing -- review \& editing: A.G., J.M., and C.K.

\clearpage

\printbibliography[heading=bibintoc]

\end{multicols}

\clearpage

\appendix
\appendixfigures

\section{Appendix}
\label{sec:appendix}

\subsection{The Full Equations of the Mean-Field Description}

In the main part, we have derived the mean-field adaptation and propagation terms separately in Eq.~\eqref{eq:mean-field-equation_adaptation} and \eqref{eq:mean-field-equation_propagation}, respectively. Through Eq.~\eqref{eq:mean-field-equation}, these are combined to give the full set of equations for a mean-field description up to order 2.

\begin{equation}
    \begin{split}
        \derivative{t} [\A]        & = (1 - p) \sum_{(m, n) \in \bar{Q}^K} [\A^m \B^n] (n \, \eta_{\A}(m, n) - m \, \eta_{\B}(m, n)) \\
        \derivative{t} [\B]        & = (1 - p) \sum_{(m, n) \in \bar{Q}^K} [\A^m \B^n] (m \, \eta_{\B}(m, n) - n \, \eta_{\A}(m, n)) \\
        \derivative{t} [\A^a \B^b] & =
        \begin{aligned}[t]
             & \mathrelphantom{+} p \Biggl(
            \begin{aligned}[t]
                 & \frac{a+1}{a+b+1} [\A^{a + 1} \B^b] \indicator{\bar{Q}^K}(a+1, b) + \frac{b + 1}{a + b + 1} [\A^a \B^{b+1}] \indicator{\bar{Q}^K}(a, b + 1) - [\A^a \B^b] \indicator{\bar{Q}^K}(a, b) \\
                 & + \sum_{(m, n) \in \bar{Q}^K} [\A^m \B^n] \left(\frac{m}{m + n} (\pi_{\A}(a - 1, b) - \pi_{\A}(a, b)) + \frac{n}{m + n} (\pi_{\B}(a, b - 1) - \pi_{\B}(a, b))\right)
                \Biggr)
            \end{aligned} \\
            %
             & + (1 - p) \left( \vphantom{\sumintersectionsAplus} \right.
            \begin{aligned}[t]
                 & - [\A^a \B^b] \indicator{\bar{Q}^K}(a, b)
                + \sum_{1 \leq \mu \leq a-1} \eta_{\A}(a - \mu, \mu) [\A^{a - \mu} \B^\mu] \delta_{b, 0}
                + \sum_{1 \leq \nu \leq b-1} \eta_{\B}(\nu, b - \nu) [\A^\nu \B^{b - \nu}] \delta_{a, 0}                                              \\[1.5ex]
                %
                 & + \sum_{(m, n) \in \bar{Q}^K} \eta_{\A}(m,n)
                \begin{aligned}[t]
                    \left( \mathrelphantom{-} \sumintersectionsAplus \right. & (1 + \delta_{m, a - \nu}\delta_{n - \nu, b})
                    [\A^{a - \mu - \nu} \B^b (\A^\mu \B^\nu) \A^{m-\mu} \B^{n-\nu}]                                         \\
                    - \sumintersectionsAminus                                & (1 + \delta_{m, a}\delta_{n, b})
                    [\A^{a - \mu} \B^{b - \nu} (\A^\mu \B^\nu) \A^{m-\mu} \B^{n-\nu}]
                    \left. \vphantom{\sumintersectionsAplus} \right)
                \end{aligned} \\
                %
                 & + \sum_{(m,n) \in \bar{Q}^K} \eta_{\B}(m, n)
                \begin{aligned}[t]
                    \left( \mathrelphantom{-} \sumintersectionsBplus \right. & (1 + \delta_{m - \mu, a} \delta_{n, b - \mu})
                    [\A^a \B^{b-\mu-\nu} (\A^\mu \B^\nu) \A^{m - \mu} \B^{n - \nu}]                                          \\
                    - \sumintersectionsBminus                                & (1 + \delta_{m, a} \delta_{n, b})
                    [\A^{a-\mu} \B^{b-\nu} (\A^\mu \B^\nu) \A^{m-\mu} \B^{n-\nu}]
                    \left. \vphantom{\sumintersectionsBplus} \right)
                    \left. \vphantom{\sumintersectionsAplus} \right)
                \end{aligned}                \\
            \end{aligned}                                                                                                    \\
        \end{aligned}
    \end{split}
\end{equation}

While the equations describing the evolution of the expected number of hyperedges above apply to active and inactive hyperedges alike, they simplify considerably in the case of the latter. For $a \geq 2$ and $b = 0$, the equations reduce to

\begin{equation}
    \begin{split}
        \derivative{t} [\A^a] & =
        \begin{aligned}[t]
             & \mathrelphantom{+} p \Biggl(
            \frac{1}{a + 1} [\A^a \B] \indicator{\bar{Q}^K}(a, 1) + \sum_{(m, n) \in \bar{Q}^K} [\A^m \B^n] \left(\frac{m}{m + n} (\pi_{\A}(a - 1, 0) - \pi_{\A}(a, 0)) - \frac{n}{m + n} \pi_{\B}(a, 0)\right)
            \Biggr)                                                                                                         \\
            %
             & + (1 - p) \left( \vphantom{\sumintersectionsAplus} \right.
            \begin{aligned}[t]
                 & \mathrelphantom{+} \sum_{1 \leq \mu \leq a-1} \eta_{\A}(a - \mu, \mu) [\A^{a - \mu} \B^\mu] \\[1.5ex]
                %
                 & + \sum_{(m, n) \in \bar{Q}^K} \eta_{\A}(m,n)
                \sum_{\triplestack{0 \leq \mu \leq m, 1 \leq \nu \leq n}{\mu + \nu < \max(m+n, a)}{\mu + \nu \leq a}}
                (1 + \delta_{m, a - \nu}\delta_{n - \nu, 0})
                [\A^{a - \mu - \nu} (\A^\mu \B^\nu) \A^{m-\mu} \B^{n-\nu}]                                   \\
                %
                 & - \sum_{(m,n) \in \bar{Q}^K} \eta_{\B}(m, n)
                \sum_{\triplestack{1 \leq \mu \leq m}{\mu < \max(m+n, a)}{\mu \leq a}}
                [\A^{a-\mu} (\A^\mu) \A^{m-\mu} \B^n]
                \left. \vphantom{\sumintersectionsAplus} \right),                                            \\
            \end{aligned} \\
        \end{aligned}
    \end{split}
\end{equation}

and, similarly, for $a = 0$ and $b \geq 2$, to

\begin{equation}
    \begin{split}
        \derivative{t} [\B^b] & =
        \begin{aligned}[t]
             & \mathrelphantom{+} p \Biggl(
            \frac{1}{b+1} [\A \B^b] \indicator{\bar{Q}^K}(1, b) + \sum_{(m, n) \in \bar{Q}^K} [\A^m \B^n] \left(- \frac{m}{m + n} \pi_{\A}(0, b) + \frac{n}{m + n} (\pi_{\B}(0, b - 1) - \pi_{\B}(0, b))\right)
            \Biggr)                                                                                                                   \\
            %
             & + (1 - p) \left( \vphantom{\sumintersectionsAplus} \right.
            \begin{aligned}[t]
                 & \mathrelphantom{+} \sum_{1 \leq \nu \leq b-1} \eta_{\B}(\nu, b - \nu) [\A^\nu \B^{b - \nu}] \\[1.5ex]
                %
                 & - \sum_{(m, n) \in \bar{Q}^K} \eta_{\A}(m,n)
                \sum_{\triplestack{1 \leq \nu \leq n}{\nu < \max(m+n, b)}{\nu \leq b}}
                [\B^{b - \nu} (\B^\nu) \A^{m} \B^{n-\nu}]                                                    \\
                %
                 & + \sum_{(m,n) \in \bar{Q}^K} \eta_{\B}(m, n)
                \sum_{\triplestack{1 \leq \mu \leq m, 0 \leq \nu \leq n}{\mu + \nu < \max(m+n, b)}{\mu + \nu \leq b}}
                (1 + \delta_{m - \mu, 0} \delta_{n, b - \mu})
                [\B^{b-\mu-\nu} (\A^\mu \B^\nu) \A^{m - \mu} \B^{n - \nu}]
                \left. \vphantom{\sumintersectionsAplus} \right).                                            \\
            \end{aligned} \\
        \end{aligned}
    \end{split}
\end{equation}

\subsection{Simulation of the Model}

We simulate the adaptive voter model on a hypergraph in discrete time. In this case, a random hyperedge is chosen at every time step. If the hyperedge is inactive, nothing happens; otherwise, the network evolves by the same rules as in the continuous model: The adaptation rule is executed with probability $p$, and the propagation rule with probability $1 - p$. To align the discrete and continuous time scales, the discrete time step is set to $\Delta t = 1/\sum_{i=2}^K M_{i}$. This choice of $\Delta t$ ensures that the frequency of updates per hyperedge per unit of time is independent of the size of the system.

To generate a hypergraph with a fixed number of hyperedges of every size, $\mathbf{M} = (M_{2}, M_{3}, \hdots,  M_{K})$, we employ a rejection sampling process. A set of vertices of a given cardinality $2 \leq i \leq K$ is drawn from the set of all vertices $V$. If all vertices are unique and if this set of vertices does not already exist in the hypergraph, a hyperedge with those vertices is created and added to the hypergraph; otherwise, the set of vertices is rejected. This process is repeated for each hyperedge size until there are exactly $M_{i}$ hyperedges of every cardinality $2 \leq i \leq K$.

After the hypergraph is generated, every vertex is assigned either a state $\A$ with probability $(1 + \tanglebr{\mathfrak{m}_{0}})/2$ or a state $\B$ with probability $(1 - \tanglebr{\mathfrak{m}_{0}})/2$ for an expected initial magnetization density $\tanglebr{\mathfrak{m}_{0}}$ with $-1 \leq \tanglebr{\mathfrak{m}_{0}} \leq 1$. Note that the actual initial magnetization $\mathfrak{m}_{0}$ in a given realization follows a Binomial distribution $B(N, q)$ with $q = (1 + \tanglebr{\mathfrak{m}_{0}})/2$.

\end{document}